\documentclass{eptcs}
\usepackage{amsmath}
\usepackage{caption}
\usepackage{subcaption}
\usepackage{minibox}
\usepackage{amssymb}
\usepackage{amsthm}
\usepackage[toc,page]{appendix}
\usepackage{graphicx}
\usepackage{dcolumn}
\usepackage{bm}
\usepackage{hyperref}
\usepackage{float}
\usepackage{tikzfig}

\tikzstyle{every picture}=[baseline=-0.25em]
\tikzstyle{dotpic}=[scale=0.6]
\tikzstyle{diredges}=[every to/.style={diredge}]
\tikzstyle{dot graph}=[shorten <=-0.1mm,shorten >=-0.1mm,scale=0.6]
\tikzstyle{plot point}=[circle,fill=black,minimum width=2mm,inner sep=0]

\tikzstyle{braceedge}=[decorate,decoration={brace,amplitude=2mm,raise=-1mm}]
\tikzstyle{small braceedge}=[decorate,decoration={brace,amplitude=1mm,raise=-1mm}]
\tikzstyle{left hook arrow}=[left hook-latex]
\tikzstyle{right hook arrow}=[right hook-latex]

\tikzstyle{dot}=[inner sep=0.7mm,minimum width=0pt,minimum height=0pt,fill=black,draw=black,shape=circle]
\tikzstyle{white dot}=[dot,fill=white]
\tikzstyle{alt white dot}=[white dot,label={[xshift=2.9mm,yshift=-0.1mm]left:$\cdot$}]
\tikzstyle{gray dot}=[dot,fill=gray!50]
\tikzstyle{box vertex}=[draw=black,rectangle]
\tikzstyle{whitebg}=[fill=white,inner sep=2pt]
\tikzstyle{graph state vertex}=[sg vertex,fill=black]

\tikzstyle{wide point}=[fill=white,draw=black,shape=isosceles triangle,shape border rotate=90,isosceles triangle stretches=true,inner sep=1pt,minimum width=1.5cm,minimum height=5mm]
\tikzstyle{wide copoint}=[fill=white,draw=black,shape=isosceles triangle,shape border rotate=-90,isosceles triangle stretches=true,inner sep=1pt,minimum width=1.5cm,minimum height=5mm]
\tikzstyle{symm}=[ultra thick,shorten <=-1mm,shorten >=-1mm]

\tikzstyle{square box}=[rectangle,fill=white,draw=black,minimum height=6mm,minimum width=6mm]
\tikzstyle{square gray box}=[rectangle,fill=gray!30,draw=black,minimum height=6mm,minimum width=6mm]
\tikzstyle{point}=[regular polygon,regular polygon sides=3,draw=black,scale=0.75,inner sep=-0.5pt,minimum width=7mm,fill=white]
\tikzstyle{copoint}=[point,regular polygon rotate=180,fill=white]
\tikzstyle{gray point}=[point,fill=gray!40!white]
\tikzstyle{gray copoint}=[copoint,fill=gray!40!white]

\newcommand{\edgearrow}{{\arrow[black]{>}}}
\newcommand{\edgetick}{{\arrow[black,scale=0.7,very thick]{|}}}
\tikzstyle{diredge}=[postaction=decorate,decoration={markings, mark=at position 0.55 with \edgearrow}]
\tikzstyle{medium diredge}=[postaction=decorate,decoration={markings, mark=at position 0.75 with \edgearrow}]
\tikzstyle{short diredge}=[->]
\tikzstyle{halfedge}=[-)]
\tikzstyle{other halfedge}=[(-]
\tikzstyle{freeedge}=[(-)]
\tikzstyle{white edge}=[line width=5pt,white]
\tikzstyle{tick}=[postaction=decorate,decoration={markings, mark=at position 0.5 with \edgetick}]
\tikzstyle{small map edge}=[|-latex, gray!60!blue, shorten <=0.9mm, shorten >=0.5mm]
\tikzstyle{thick dashed edge}=[very thick,dashed,gray!40]
\tikzstyle{map edge}=[|-latex,very thick, gray!40, shorten <=1mm, shorten >=0.5mm]
\tikzstyle{tickedge}=[postaction=decorate,
  decoration={markings, mark=at position 0.5 with \edgetick}]
\tikzstyle{dirtickedge}=[postaction=decorate,
  decoration={markings, mark=at position 0.5 with \edgetick},
  decoration={markings, mark=at position 0.85 with \edgearrow}]
\tikzstyle{dirdoubletickedge}=[postaction=decorate,
  decoration={markings, mark=at position 0.4 with \edgetick},
  decoration={markings, mark=at position 0.6 with \edgetick},
  decoration={markings, mark=at position 0.85 with \edgearrow}]

\tikzstyle{arrs}=[-latex,font=\small,auto]
\tikzstyle{arrow plain}=[arrs]
\tikzstyle{arrow dashed}=[dashed,arrs]
\tikzstyle{arrow bold}=[very thick,arrs]
\tikzstyle{arrow hide}=[draw=white!0,-]
\tikzstyle{arrow reverse}=[latex-]
\tikzstyle{cdnode}=[]

\tikzstyle{cnot}=[fill=white,shape=circle,inner sep=-1.4pt]
\tikzstyle{wire label}=[font=\footnotesize, auto]

\floatstyle{boxed}

\usepackage{xy}
\xyoption{matrix}
\xyoption{frame}
\xyoption{arrow}
\xyoption{arc}

\usepackage{ifpdf}
\ifpdf
\else
\PackageWarningNoLine{Qcircuit}{Qcircuit is loading in Postscript mode.  The Xy-pic options ps and dvips will be loaded.  If you wish to use other Postscript drivers for Xy-pic, you must modify the code in Qcircuit.tex}
\xyoption{ps}
\xyoption{dvips}
\fi

\entrymodifiers={!C\entrybox}

\newcommand{\bra}[1]{\ensuremath{\left\langle#1\right|}}
\newcommand{\ket}[1]{\ensuremath{\left|#1\right\rangle}}

\usepackage{graphicx}
\usepackage{dcolumn}
\usepackage{bm}

\pgfdeclarelayer{edgelayer}
\pgfdeclarelayer{nodelayer}
\pgfsetlayers{edgelayer,nodelayer,main}

\tikzstyle{none}=[inner sep=0pt]
\definecolor{hexcolor0xff0000}{rgb}{1.000,0.000,0.000}
\definecolor{hexcolor0x000000}{rgb}{0.000,0.000,0.000}
\definecolor{hexcolor0xffff00}{rgb}{1.000,1.000,0.000}
\definecolor{Lime}{rgb}{0.000,1.000,0.000}
\definecolor{hexcolor0x030000}{rgb}{0.012,0.000,0.000}

\tikzstyle{rn}=[circle,fill=hexcolor0xff0000,draw=hexcolor0x000000,line width=0.8 pt, minimum width=8pt, minimum height=8pt, inner sep=0.4pt]
\tikzstyle{gn}=[circle,fill=Lime,draw=hexcolor0x000000,line width=0.8 pt, minimum width=8pt, minimum height=8pt, inner sep=0.4pt]
\tikzstyle{yn}=[circle,fill=hexcolor0xffff00,draw=hexcolor0x000000,line width=0.8 pt, minimum width=8pt, minimum height=8pt, inner sep=0.4pt]
\tikzstyle{diam}=[shape=diamond,fill=hexcolor0x030000,draw=hexcolor0x000000,minimum width=8pt, minimum height=8pt, inner sep=0pt]
\tikzstyle{Had}=[rectangle,fill=hexcolor0xffff00,draw=hexcolor0x000000,minimum width=8pt, minimum height=8pt, inner sep=0.2pt]
\tikzstyle{rpoint}=[up triangle,fill=hexcolor0xff0000,draw=hexcolor0x000000,minimum width=8pt, minimum height=8pt, inner sep=0.2pt]
\tikzstyle{gpoint}=[up triangle,fill=Lime,draw=hexcolor0x000000,line width=0.8 pt, minimum width=8pt, minimum height=8pt, inner sep=0.2pt]

\tikzstyle{simple}=[-,draw=hexcolor0x000000,line width=2.000]
\tikzstyle{arrow}=[-,draw=hexcolor0x000000,postaction={decorate},decoration={markings,mark=at position .5 with {\arrow{>}}},line width=2.000]
\tikzstyle{tick}=[-,draw=hexcolor0x000000,postaction={decorate},decoration={markings,mark=at position .5 with {\draw (0,-0.1) -- (0,0.1);}},line width=2.000]
\begin{document}

\title{Depicting qudit quantum mechanics and mutually unbiased qudit theories}
\author{Andr$\acute{e}$ Ranchin
\institute{University of Oxford, Department of Computer Science, Quantum Group}
\institute{Imperial College London, Department of Physics, Controlled Quantum Dynamics}}

\def\titlerunning{Qudit ZX calculus and MUQT}
\def\authorrunning{Andr$\acute{e}$ Ranchin}

\maketitle

\begin{abstract}

We generalize the ZX calculus to quantum systems of dimension higher than two. The resulting calculus is sound and universal for quantum mechanics. We define the notion of a mutually unbiased qudit theory and study two particular instances of these theories in detail: \textit{qudit stabilizer quantum mechanics} and \textit{Spekkens-Schreiber toy theory for dits}. The calculus allows us to analyze the structure of qudit stabilizer quantum mechanics and provides a geometrical picture of qudit stabilizer theory using D-toruses, which generalizes the Bloch sphere picture for qubit stabilizer quantum mechanics. We also use our framework to describe generalizations of Spekkens toy theory to higher dimensional systems. This gives a novel proof that qudit stabilizer quantum mechanics and Spekkens-Schreiber toy theory for dits are operationally equivalent in three dimensions. The qudit pictorial calculus is a useful tool to study quantum foundations, understand the relationship between qubit and qudit quantum mechanics, and provide a novel, high level description of quantum information protocols.   

\end{abstract}

\maketitle

\section{Introduction}

An interesting approach to understanding the foundations of quantum mechanics is to study sets of alternative theories which exhibit similar structural or physical features as quantum theory. Several mathematical formalisms for operational physical theories have been proposed \cite{Ab04,Ba07} which encompass quantum mechanics as one possible theory within a space of different potential theories. These provide a setting in which we can determine which features  are truly particular to quantum theory and which ones are more generic. This approach can pave the way towards novel axiomatizations of quantum mechanics and could yield precious clues about future physical theories which may supersede quantum theory, such as a theory of quantum gravity. As Lewis Carroll aptly put it: ``If you don't know where you are going, any road will get you there".

Symmetric monoidal categories (SMCs) provide a general framework for physical theories since they contain two interacting modes, $\otimes$ and $\circ$, of composing systems and processes. Previous work has investigated which additional structure must be imposed on a SMC in order to recover the structure of quantum theory \cite{Ab04}. This approach has yielded an intuitive graphical language, called the \textbf{ZX calculus}, which allows us to explicitly formulate quantum mechanics within a symmetric monoidal category \cite{Du11}. 

More precisely, the ZX calculus is a two coloured pictorial calculus for qubits whose diagrams are generated by composing basic process diagrams and which has a set of rule equations specifying how one diagram can be transformed into another. The calculus is sound and universal for quantum mechanics and it has been shown that it is complete for stabilizer quantum mechanics, given a certain choice of phases \cite{Back12}. The ZX calculus has proven useful in the study of quantum foundations \cite{Co12}, quantum computation \cite{Dun10} and quantum error-correction \cite{Hors11}. 

In the present article, we generalize the ZX calculus to qudit systems and show that the resulting calculus is universal for quantum mechanics. We anticipate that the new calculus will provide a practical tool to study quantum information and computation from a high-level point of view. For example, the qudit calculus for dimensions higher than two should be well suited to understanding structural properties of quantum algorithms, quantum key distribution and quantum error-correction. Moreover, as the complexity of the quantum systems we study will grow, computer software such as Quantomatic \cite{Kiss09}, which allows automated reasoning within the calculus, may play an important role in the design of future quantum networks. 

For the time being, we focus on using key ideas from the ZX qudit calculus to study quantum foundations. In particular, we define the notion of a \textbf{mutually unbiased qudit theory} (MUQT), which corresponds to a symmetric monoidal category whose observable structures are all mutually unbiased. These MUQTs can be classified in terms of a particular Abelian group, called the \textbf{phase group}. 

Previous work has shown that in the case of qubits \cite{Ed11}, there are essentially two MUQTs: stabilizer quantum mechanics \cite{Got97}, which has phase group $\mathbb{Z}_4$, and Spekken's toy theory for bits \cite{Spek07}, which has phase group $\mathbb{Z}_2 \times \mathbb{Z}_2 $. Furthermore, the phase groups of these theories determine whether or not they admit a local hidden variable model. We aim to generalize this work to higher dimensional systems.  

This article focuses on the study of two interesting families of MUQTs, corresponding to \textit{stabilizer quantum theory for qudits} \cite{Gott98} and \textit{Spekkens-Schreiber's toy theory for dits} \cite{Scsp12}. This is a first step towards a full classification of MUQTs and a thorough study of the relationship between physical features of these theories and the properties of their phase groups.

\section{The ZX calculus for qudit quantum mechanics:}

We now present the ZX calculus for qudit quantum mechanics. This is a generalization of the standard qubit ZX calculus \cite{Du08}. The mathematical background upon which the calculus is built is presented in Appendix A. We briefly repeat a few essential definitions. An observable structure, which is a generalization of the Hilbert space concept of an orthonormal basis, consists of a copying map $\delta:$ \begin{tikzpicture}[scale=0.5]
	\begin{pgfonlayer}{nodelayer}
		\node [style=none] (0) at (0, 1) {};
		\node [style=none] (1) at (0.4999993, -1) {};
		\node [style=none] (2) at (-0.4999993, -1) {};
		\node [style=gn] (3) at (0, -0) {};
	\end{pgfonlayer}
	\begin{pgfonlayer}{edgelayer}
		\draw (0.center) to (3);
		\draw (3) to (1.center);
		\draw (3) to (2.center);
	\end{pgfonlayer}
\end{tikzpicture} and a deleting map  
$\epsilon:$ \begin{tikzpicture}[scale=0.5]
	\begin{pgfonlayer}{nodelayer}
		\node [style=none] (0) at (0, 0.5) {};
		\node [style=gn] (1) at (0, -0.5) {};
	\end{pgfonlayer}
	\begin{pgfonlayer}{edgelayer}
		\draw (0.center) to (1);
	\end{pgfonlayer}
\end{tikzpicture} satisfying certain algebraic conditions. A state (or point) $\psi$ is \textit{classical} (or an eigenstate) for an observable structure if it is copied by the copying map and deleted by the deleting map. $\psi$ is \textit{unbiased} with respect to an observable structure if: $s(\delta^{\dagger} \circ( \psi \otimes \psi^{\star}))= \epsilon^{\dagger}$ for some scalar s. 

Given an observable structure, each state $\psi$ has a corresponding \textit{phase map}: $\Lambda(\psi):=\delta^{\dagger} \circ(\psi \otimes \mathbb{I})$. The set of all phase maps corresponding to unbiased states for an observable structure, together with map composition, form a group called the \textbf{phase group}. 
We will now present the rules of the calculus and its relationship to quantum theory.

General network diagrams are built out of parallel (tensor product) and downward compositions of generating diagrams from Figure \ref{ZXgen}.

\begin{figure}[H]
\begin{center}
\frame{\begin{tikzpicture}[scale=0.6]
	\path [use as bounding box] (-10,-2) rectangle (14,2);
	\begin{pgfonlayer}{nodelayer}
		\node [style=none] (0) at (-9, 1.25) {};
		\node [style=none] (1) at (-9, -1.25) {};
		\node [style=none] (2) at (-7.75, 1.25) {};
		\node [style=none] (3) at (-6.25, 1.25) {};
		\node [style=none] (4) at (-7.75, -1.25) {};
		\node [style=none] (5) at (-6.25, -1.25) {};
		\node [style=none] (6) at (-5, 1.25) {};
		\node [style=none] (7) at (-5, -1.25) {};
		\node [style=Had] (8) at (-5, -0) {F};
		\node [style=rn] (9) at (0.5000002, -0) {};
		\node [style=none] (10) at (-0.9999997, 1.25) {};
		\node [style=none] (11) at (-0.2499997, 1.25) {};
		\node [style=none] (12) at (0.5000002, 1.25) {...};
		\node [style=none] (13) at (1.25, 1.25) {};
		\node [style=none] (14) at (1.999999, 1.25) {};
		\node [style=none] (15) at (-0.2499997, -1.25) {};
		\node [style=none] (16) at (1.25, -1.25) {};
		\node [style=none] (17) at (-0.9999997, -1.25) {};
		\node [style=none] (18) at (1.999999, -1.25) {};
		\node [style=none] (19) at (0.5000002, -1.25) {...};
		\node [style=none] (20) at (6.25, 1.25) {...};
		\node [style=none] (21) at (4.75, -1.25) {};
		\node [style=none] (22) at (4.75, 1.25) {};
		\node [style=none] (23) at (7.75, -1.25) {};
		\node [style=none] (24) at (7, -1.25) {};
		\node [style=none] (25) at (5.5, 1.25) {};
		\node [style=none] (26) at (7.75, 1.25) {};
		\node [style=none] (27) at (5.5, -1.25) {};
		\node [style=none] (28) at (7, 1.25) {};
		\node [style=none] (29) at (6.25, -1.25) {...};
		\node [style=gn] (30) at (6.25, -0) {};
		\node [style=none] (31) at (-8.25, -0) {;};
		\node [style=none] (32) at (-6, -0) {;};
		\node [style=none] (33) at (4.25, -0) {;};
		\node [style=none] (34) at (-4, -0) {;};
		\node [style=none] (35) at (-1.5, -0) {;};
		\node [style=Had] (36) at (-3, -0) {$F^{\dagger}$};
		\node [style=none] (37) at (-3, 1.25) {};
		\node [style=none] (38) at (-3, -1.25) {};
		\node [style=none] (39) at (2.5, -0) {$\alpha_1$, ..., $\alpha_{D-1}$};
		\node [style=none] (40) at (8.25, -0) {$\beta_1$, ..., $\beta_{D-1}$};
		\node [style=none] (41) at (10, -0) {;};
		\node [style=none] (42) at (12, -0.7499999) {$s \in \mathbb{C}$};
		\node [style=none] (43) at (12, 0.7499999) {scalar terms};
	\end{pgfonlayer}
	\begin{pgfonlayer}{edgelayer}
		\draw (0.center) to (1.center);
		\draw (2.center) to (5.center);
		\draw (3.center) to (4.center);
		\draw (6.center) to (8);
		\draw (8) to (7.center);
		\draw (9) to (10.center);
		\draw (9) to (11.center);
		\draw (9) to (13.center);
		\draw (9) to (14.center);
		\draw (9) to (17.center);
		\draw (9) to (15.center);
		\draw (9) to (16.center);
		\draw (9) to (18.center);
		\draw (22.center) to (30);
		\draw (25.center) to (30);
		\draw (28.center) to (30);
		\draw (26.center) to (30);
		\draw (30) to (21.center);
		\draw (30) to (27.center);
		\draw (30) to (24.center);
		\draw (30) to (23.center);
		\draw (37.center) to (36);
		\draw (36) to (38.center);
	\end{pgfonlayer}
\end{tikzpicture}}
\end{center}
\caption{Generating diagrams for the ZX network.}
\label{ZXgen}
\end{figure}

The rules of the \textbf{\textit{qudit ZX calculus}} are the (S), (D), (B), (K), and (F) rules below (and their reversed colour counterparts), together with a (T) rule which states that after identifying the inputs and outputs of any part of a ZX network, any topological deformation of the internal structure does not matter. 

\begin{center}
\begin{tikzpicture}[scale=0.55]
	\path [use as bounding box] (-14.75,-13) rectangle (14.5,17);
	\begin{pgfonlayer}{nodelayer}
		\node [style=none] (0) at (2.750001, 5) {};
		\node [style=none] (1) at (2.750001, 2) {};
		\node [style=none] (2) at (7.75, 2) {};
		\node [style=none] (3) at (3.750001, 5) {};
		\node [style=none] (4) at (3.750001, 2) {};
		\node [style=gn] (5) at (7.25, 3) {};
		\node [style=gn] (6) at (2.750001, 4) {};
		\node [style=none] (7) at (5.25, 3.5) {=};
		\node [style=rn] (8) at (7.25, 4) {};
		\node [style=gn] (9) at (3.750001, 4) {};
		\node [style=none] (10) at (6.75, 5) {};
		\node [style=none] (11) at (13, 3.5) {(B2)};
		\node [style=rn] (12) at (2.750001, 3) {};
		\node [style=none] (13) at (7.75, 5) {};
		\node [style=rn] (14) at (3.750001, 3) {};
		\node [style=none] (15) at (6.75, 2) {};
		\node [style=none] (16) at (-5.25, -0.4999997) {(K1)};
		\node [style=none] (17) at (-6.000001, 15.75) {};
		\node [style=none] (18) at (0, 15.75) {};
		\node [style=none] (19) at (0.9999989, 11.75) {...};
		\node [style=rn] (20) at (0.9999989, 13.75) {};
		\node [style=none] (21) at (-5.25, 8.25) {(S2)};
		\node [style=none] (22) at (-10.25, 11.75) {...};
		\node [style=none] (23) at (-11, 11.75) {};
		\node [style=none] (24) at (-2.250001, 13.75) {=};
		\node [style=none] (25) at (-8.499999, 13.75) {...};
		\node [style=none] (26) at (13, 14) {(S1)};
		\node [style=none] (27) at (-8.499999, 15.75) {};
		\node [style=none] (28) at (-14, 6.75) {};
		\node [style=none] (29) at (-9.500001, 15.75) {...};
		\node [style=none] (30) at (2, 11.75) {};
		\node [style=gn] (31) at (-14, 8.25) {};
		\node [style=rn] (32) at (-14, 8.25) {};
		\node [style=none] (33) at (-6.75, 15.75) {...};
		\node [style=none] (34) at (-14, 9.75) {};
		\node [style=none] (35) at (2, 15.75) {};
		\node [style=none] (36) at (-7, 6.75) {};
		\node [style=none] (37) at (-8.499999, 11.5) {};
		\node [style=none] (38) at (-7, 9.75) {};
		\node [style=none] (39) at (-8.000001, 8.25) {=};
		\node [style=none] (40) at (-10.5, 15.75) {};
		\node [style=none] (41) at (0.9999989, 15.75) {...};
		\node [style=rn] (42) at (-9.500001, 14.75) {};
		\node [style=none] (43) at (-7.5, 15.75) {};
		\node [style=none] (44) at (-6.499999, 11.5) {};
		\node [style=none] (45) at (0, 11.75) {};
		\node [style=none] (46) at (-7.5, 11.75) {...};
		\node [style=none] (47) at (-9.500001, 11.75) {};
		\node [style=none] (48) at (-9.750001, 3.5) {=};
		\node [style=none] (49) at (-7.25, 3) {};
		\node [style=rn] (50) at (-7.25, 4) {};
		\node [style=none] (51) at (-5.25, 3.5) {(B1)};
		\node [style=none] (52) at (-8.249999, 3) {};
		\node [style=rn] (53) at (-8.249999, 4) {};
		\node [style=none] (54) at (-10.75, 2.5) {};
		\node [style=gn] (55) at (-11.75, 3.5) {};
		\node [style=rn] (56) at (-11.75, 4.5) {};
		\node [style=none] (57) at (-12.75, 2.5) {};
		\node [style=rn] (58) at (-8.750001, -1) {k};
		\node [style=gn] (59) at (-13, -1) {};
		\node [style=rn] (60) at (-13, -0) {k};
		\node [style=none] (61) at (-8.750001, -2) {};
		\node [style=none] (62) at (-9.75, -2) {};
		\node [style=none] (63) at (-14, -2) {};
		\node [style=none] (64) at (-11, -0.4999997) {=};
		\node [style=rn] (65) at (-9.75, -1) {k};
		\node [style=none] (66) at (-12, -2) {};
		\node [style=none] (67) at (6.499999, 13.75) {$\alpha_1+\beta_1$, $\alpha_2+ \beta_2$,  ..., $\alpha_{D-1}+\beta_{D-1}$};
		\node [style=rn] (68) at (-7.5, 12.5) {};
		\node [style=none] (69) at (-12.75, 14.75) {$\alpha_1$, $\alpha_2$, ..., $\alpha_{D-1}$};
		\node [style=none] (70) at (-4.499999, 12.5) {$\beta_1$, $\beta_2$, ..., $\beta_{D-1}$};
		\node [style=none] (71) at (-11.25, 9.75) {};
		\node [style=none] (72) at (-11.25, 6.5) {};
		\node [style=none] (73) at (-12.75, 8.25) {:=};
		\node [style=rn] (74) at (-11.25, 8.25) {};
		\node [style=none] (75) at (-13, 1) {};
		\node [style=none] (76) at (-9.249999, 1) {};
		\node [style=gn] (77) at (-9.249999, -0) {};
		\node [style=Had] (78) at (-2.250001, -10.25) {$F$};
		\node [style=none] (79) at (2.749999, -4.5) {};
		\node [style=none] (80) at (-10.25, -8.499999) {};
		\node [style=Had] (81) at (-7.750001, -7.5) {F};
		\node [style=rn] (82) at (-8.499999, -6.5) {};
		\node [style=Had] (83) at (-2.250001, -11.25) {$F^{\dagger}$};
		\node [style=none] (84) at (2.749999, -8.499999) {};
		\node [style=none] (85) at (2, -4.5) {...};
		\node [style=Had] (86) at (-6.75, -7.5) {F};
		\node [style=Had] (87) at (-9.250001, -5.5) {$F^{\dagger}$};
		\node [style=none] (88) at (-8.499999, -4.5) {...};
		\node [style=none] (89) at (-5.5, -6.5) {$\alpha_1,\alpha_2, ..., \alpha_{D-1} $};
		\node [style=Had] (90) at (-5.25, -10.25) {$F^{\dagger}$};
		\node [style=none] (91) at (0.5000007, -4.5) {};
		\node [style=none] (92) at (-10.25, -4.5) {};
		\node [style=none] (93) at (-8.499999, -8.499999) {...};
		\node [style=none] (94) at (0.7500011, -9.5) {};
		\node [style=none] (95) at (1.249999, -4.5) {};
		\node [style=none] (96) at (-9.500001, -4.5) {};
		\node [style=Had] (97) at (-7.5, -5.5) {$F^{\dagger}$};
		\node [style=none] (98) at (-2.250001, -9.5) {};
		\node [style=none] (99) at (3.5, -4.5) {};
		\node [style=none] (100) at (-7.25, -4.5) {};
		\node [style=gn] (101) at (-8.499999, -6.5) {};
		\node [style=none] (102) at (-5.25, -12) {};
		\node [style=none] (103) at (3.5, -8.499999) {};
		\node [style=none] (104) at (-6.499999, -4.5) {};
		\node [style=Had] (105) at (-6.499999, -5.5) {$F^{\dagger}$};
		\node [style=none] (106) at (-2.250001, -12) {};
		\node [style=none] (107) at (-3.75, -10.75) {=};
		\node [style=rn] (108) at (2, -6.5) {};
		\node [style=none] (109) at (1.249999, -8.499999) {};
		\node [style=none] (110) at (-6.75, -8.499999) {};
		\node [style=none] (111) at (-2.250001, -6.5) {=};
		\node [style=Had] (112) at (-5.25, -11.25) {$F$};
		\node [style=none] (113) at (13, -10.75) {(F2)};
		\node [style=none] (114) at (2, -8.499999) {...};
		\node [style=none] (115) at (-7.5, -8.499999) {};
		\node [style=Had] (116) at (-10.25, -7.5) {F};
		\node [style=none] (117) at (-0.7500011, -10.75) {=};
		\node [style=none] (118) at (13, -6.5) {(F1)};
		\node [style=none] (119) at (-5.25, -9.5) {};
		\node [style=none] (120) at (4.75, -6.5) {$\alpha_1,\alpha_2, ..., \alpha_{D-1} $};
		\node [style=none] (121) at (0.5000007, -8.499999) {};
		\node [style=none] (122) at (-9.500001, -8.499999) {};
		\node [style=Had] (123) at (-9.250001, -7.5) {F};
		\node [style=Had] (124) at (-10.25, -5.5) {$F^{\dagger}$};
		\node [style=none] (125) at (0.7500011, -12) {};
		\node [style=none] (126) at (13, -0.5000004) {(K2)};
		\node [style=none] (127) at (7, -1.25) {$Neg(\alpha_1, ..., \alpha_{D-1})$  };
		\node [style=rn] (128) at (-2.249999, -1.25) {};
		\node [style=none] (129) at (-2.249999, 1.25) {};
		\node [style=none] (130) at (4.749999, 0.2500004) {k};
		\node [style=none] (131) at (4.000002, 1.25) {};
		\node [style=none] (132) at (0.5000008, 0.2500004) {$\alpha_1,\alpha_2,...,\alpha_{D-1}$};
		\node [style=gn] (133) at (4.000002, -1.25) {};
		\node [style=none] (134) at (-2.249999, -2.25) {};
		\node [style=rn] (135) at (4.000002, 0.2500004) {};
		\node [style=none] (136) at (-1.5, -1.25) {k};
		\node [style=none] (137) at (2.75, -0.4999997) {=};
		\node [style=none] (138) at (4.000002, -2.25) {};
		\node [style=gn] (139) at (-2.249999, 0.2500004) {};
		\node [style=rn] (140) at (-14, 4) {};
		\node [style=gn] (141) at (-14, 3) {};
		\node [style=gn] (142) at (1.5, 4) {};
		\node [style=rn] (143) at (1.5, 3) {};
		\node [style=none] (144) at (-9.500001, 8.25) {0, ..., 0};
		\node [style=none] (145) at (8.000001, 8) {=};
		\node [style=none] (146) at (13, 8) {(D)};
		\node [style=gn] (147) at (9, 9) {};
		\node [style=gn] (148) at (7, 7) {};
		\node [style=none] (149) at (11, 8) {$\sqrt{D}$};
		\node [style=gn] (150) at (4, 9) {};
		\node [style=none] (151) at (0, 8) {};
		\node [style=none] (152) at (3, 8) {=};
		\node [style=rn] (153) at (4, 7) {};
		\node [style=none] (154) at (10, 8) {=};
		\node [style=rn] (155) at (9, 7) {};
		\node [style=none] (156) at (6, 8) {and};
		\node [style=none] (157) at (2, 8) {};
		\node [style=rn] (158) at (7, 9) {};
		\node [style=gn] (159) at (5, 7) {};
		\node [style=rn] (160) at (5, 9) {};
		\node [style=none] (161) at (0.9999995, 9) {};
		\node [style=none] (162) at (0.9999995, 9) {};
		\node [style=rn] (163) at (0.9999995, 9.75) {};
		\node [style=rn] (164) at (0.9999995, 9) {};
		\node [style=none] (165) at (0.9999995, 7) {};
		\node [style=rn] (166) at (0.9999995, 7) {};
		\node [style=rn] (167) at (0.9999995, 6.25) {};
	\end{pgfonlayer}
	\begin{pgfonlayer}{edgelayer}
		\draw (0.center) to (6);
		\draw (3.center) to (9);
		\draw (9) to (12);
		\draw (6) to (14);
		\draw (14) to (4.center);
		\draw (12) to (1.center);
		\draw (6) to (12);
		\draw (14) to (9);
		\draw (10.center) to (8);
		\draw (13.center) to (8);
		\draw (8) to (5);
		\draw (5) to (15.center);
		\draw (5) to (2.center);
		\draw [style=none] (18.center) to (20);
		\draw [style=none] (20) to (35.center);
		\draw [style=none] (20) to (30.center);
		\draw [style=none] (20) to (45.center);
		\draw [style=none] (34.center) to (32);
		\draw [style=none] (32) to (28.center);
		\draw [style=none] (38.center) to (36.center);
		\draw [style=none, bend right=15, looseness=1.00] (42) to (47.center);
		\draw (27.center) to (42);
		\draw [style=none] (42) to (23.center);
		\draw (40.center) to (42);
		\draw (56) to (55);
		\draw (55) to (54.center);
		\draw (55) to (57.center);
		\draw (53) to (52.center);
		\draw (50) to (49.center);
		\draw (60) to (59);
		\draw (59) to (66.center);
		\draw (59) to (63.center);
		\draw (65) to (62.center);
		\draw (58) to (61.center);
		\draw [style=none] (68) to (37.center);
		\draw [style=none] (68) to (17.center);
		\draw [style=none, bend right=45, looseness=1.50] (42) to (68);
		\draw [style=none] (68) to (44.center);
		\draw [style=none, bend right=15, looseness=1.00] (68) to (43.center);
		\draw [style=none, bend left=45, looseness=1.25] (42) to (68);
		\draw (71.center) to (74);
		\draw (74) to (72.center);
		\draw (75.center) to (60);
		\draw (76.center) to (77);
		\draw (77) to (65);
		\draw (77) to (58);
		\draw [style=none] (94.center) to (125.center);
		\draw (119.center) to (90);
		\draw (90) to (112);
		\draw (112) to (102.center);
		\draw (98.center) to (78);
		\draw (78) to (83);
		\draw (83) to (106.center);
		\draw (82) to (97);
		\draw (82) to (105);
		\draw (105) to (104.center);
		\draw (97) to (100.center);
		\draw (96.center) to (87);
		\draw (87) to (82);
		\draw (82) to (124);
		\draw (124) to (92.center);
		\draw (82) to (116);
		\draw (116) to (80.center);
		\draw (82) to (123);
		\draw (123) to (122.center);
		\draw (82) to (81);
		\draw (81) to (115.center);
		\draw (82) to (86);
		\draw (86) to (110.center);
		\draw (91.center) to (108);
		\draw (95.center) to (108);
		\draw (79.center) to (108);
		\draw (108) to (99.center);
		\draw (108) to (121.center);
		\draw (109.center) to (108);
		\draw (108) to (84.center);
		\draw (108) to (103.center);
		\draw (129.center) to (139);
		\draw (139) to (128);
		\draw (128) to (134.center);
		\draw (131.center) to (135);
		\draw (135) to (133);
		\draw (133) to (138.center);
		\draw (140) to (141);
		\draw (142) to (143);
		\draw (150) to (153);
		\draw (160) to (159);
		\draw (158) to (148);
		\draw (147) to (155);
		\draw [bend left=45, looseness=1.25] (161.center) to (157.center);
		\draw [in=180, out=90, looseness=1.25] (151.center) to (161.center);
		\draw (163) to (161.center);
		\draw [in=-90, out=180, looseness=1.25] (165.center) to (151.center);
		\draw [bend left=45, looseness=1.00] (157.center) to (165.center);
		\draw (165.center) to (167);
	\end{pgfonlayer}
\end{tikzpicture}
\end{center}
where $Neg(\alpha_1, ..., \alpha_{D-1}):= \alpha_{k+1}-\alpha_{k},\alpha_{k+2}-\alpha_{k}, ...,\alpha_{D-1}-\alpha_k, -\alpha_k,\alpha_1-\alpha_{k},..., \alpha_{k-1}-\alpha_{k}$, and where there are D-1 different red k vertices which have phases $\alpha_1, ..., \alpha_{D-1}$ such that
\begin{tikzpicture}[scale=0.7]
	\begin{pgfonlayer}{nodelayer}
		\node [style=none] (0) at (0, 0.75) {};
		\node [style=rn] (1) at (0, -0) {};
		\node [style=none] (2) at (0, -0.75) {};
		\node [style=none] (3) at (0.7, -0) {k};
	\end{pgfonlayer}
	\begin{pgfonlayer}{edgelayer}
		\draw (1) to (2.center);
		\draw (0.center) to (1);
	\end{pgfonlayer}
\end{tikzpicture}
\hspace{3pt} are the phase maps corresponding to the D-1 classical points for Z whose phases are not all zero. In higher dimensions, the (K) rules give rise to more intricate interference phenomena, since the D classical points of an observable structure each permute the phase group elements. 

Diagrammatic reasoning in the qudit calculus is identical to reasoning in the qubit calculus. As before, two network diagrams can be shown to be equal by locally replacing some part of a diagram with a diagram equal to it. 

Note that the restricted case of the ZX calculus for qutrits has been studied independently (and synchronously) in \cite{Xia13}.  

As with the qubit case, we can model the calculus in Hilbert space. We interpret all diagram edges by $\mathbb{C}^D$ and elements of the qudit calculus correspond to the following Hilbert space elements: 

\begin{figure}[H]
\begin{tikzpicture}[scale=0.4]
	\begin{pgfonlayer}{nodelayer}
		\node [style=none] (0) at (0, 0.75) {};
		\node [style=none] (1) at (0, -0.75) {};
		\node [style=none] (2) at (1, 1) {};
		\node [style=none] (3) at (0.75, 1) {};
		\node [style=none] (4) at (0.5, 1) {};
		\node [style=none] (5) at (0.5, -1) {};
		\node [style=none] (6) at (0.75, -1) {};
		\node [style=none] (7) at (1, -1) {};
		\node [style=none] (8) at (-0.5, -1) {};
		\node [style=none] (9) at (-0.75, -1) {};
		\node [style=none] (10) at (-1, -1) {};
		\node [style=none] (11) at (-1, 1) {};
		\node [style=none] (12) at (-0.75, 1) {};
		\node [style=none] (13) at (-0.5, 1) {};
	\end{pgfonlayer}
	\begin{pgfonlayer}{edgelayer}
		\draw (13.center) to (11.center);
		\draw (8.center) to (10.center);
		\draw (10.center) to (11.center);
		\draw (12.center) to (9.center);
		\draw (5.center) to (7.center);
		\draw (6.center) to (3.center);
		\draw (4.center) to (2.center);
		\draw (2.center) to (7.center);
		\draw (0.center) to (1.center);
	\end{pgfonlayer}
\end{tikzpicture} = $\mathbb{I}_{D\times D}:= \sum_{k=0}^{D-1}\ket{k}\bra{k}$ \hspace{10pt} ; \hspace{10pt}
\begin{tikzpicture}[scale=0.4]
	\begin{pgfonlayer}{nodelayer}
		\node [style=none] (0) at (0.5, -1) {};
		\node [style=none] (1) at (-0.5, -1) {};
		\node [style=none] (2) at (-0.5, 1) {};
		\node [style=none] (3) at (0.5, 1) {};
		\node [style=none] (4) at (-1.5, 1.5) {};
		\node [style=none] (5) at (-1.25, 1.5) {};
		\node [style=none] (6) at (-1, 1.5) {};
		\node [style=none] (7) at (1, 1.5) {};
		\node [style=none] (8) at (1.25, 1.5) {};
		\node [style=none] (9) at (1.5, 1.5) {};
		\node [style=none] (10) at (1.25, -1.5) {};
		\node [style=none] (11) at (1, -1.5) {};
		\node [style=none] (12) at (1.5, -1.5) {};
		\node [style=none] (13) at (-1.25, -1.5) {};
		\node [style=none] (14) at (-1.5, -1.5) {};
		\node [style=none] (15) at (-1, -1.5) {};
	\end{pgfonlayer}
	\begin{pgfonlayer}{edgelayer}
		\draw (4.center) to (14.center);
		\draw (5.center) to (13.center);
		\draw (14.center) to (15.center);
		\draw (6.center) to (4.center);
		\draw (9.center) to (7.center);
		\draw (8.center) to (10.center);
		\draw (11.center) to (12.center);
		\draw (12.center) to (9.center);
		\draw (2.center) to (0.center);
		\draw (3.center) to (1.center);
	\end{pgfonlayer}
\end{tikzpicture} $=SWAP_{a,b}:=\sum_{j,k=0}^{D-1}  \ket{k}\bra{j}_a \otimes \ket{j}\bra{k}_b $

\begin{tikzpicture}[scale=0.4]
	\begin{pgfonlayer}{nodelayer}
		\node [style=none] (0) at (0, 0.75) {};
		\node [style=none] (1) at (0, -0.75) {};
		\node [style=none] (2) at (1, 1) {};
		\node [style=none] (3) at (0.75, 1) {};
		\node [style=none] (4) at (0.5, 1) {};
		\node [style=none] (5) at (0.5, -1) {};
		\node [style=none] (6) at (0.75, -1) {};
		\node [style=none] (7) at (1, -1) {};
		\node [style=none] (8) at (-0.5, -1) {};
		\node [style=none] (9) at (-0.75, -1) {};
		\node [style=none] (10) at (-1, -1) {};
		\node [style=none] (11) at (-1, 1) {};
		\node [style=none] (12) at (-0.75, 1) {};
		\node [style=none] (13) at (-0.5, 1) {};
		\node [style=Had] (14) at (0, -0) {F};
	\end{pgfonlayer}
	\begin{pgfonlayer}{edgelayer}
		\draw (13.center) to (11.center);
		\draw (8.center) to (10.center);
		\draw (10.center) to (11.center);
		\draw (12.center) to (9.center);
		\draw (5.center) to (7.center);
		\draw (6.center) to (3.center);
		\draw (4.center) to (2.center);
		\draw (2.center) to (7.center);
		\draw (0.center) to (14);
		\draw (14) to (1.center);
	\end{pgfonlayer}
\end{tikzpicture} = $Fourier$ :=$\frac{1}{\sqrt{D}}\sum_{j,k=0}^{D-1}\eta^{jk}\ket{j}\bra{k}$ \hspace{10pt} ; \hspace{10pt} 
\begin{tikzpicture}[scale=0.4]
	\begin{pgfonlayer}{nodelayer}
		\node [style=none] (0) at (0, 0.75) {};
		\node [style=none] (1) at (0, -0.75) {};
		\node [style=none] (2) at (1, 1) {};
		\node [style=none] (3) at (0.75, 1) {};
		\node [style=none] (4) at (0.5, 1) {};
		\node [style=none] (5) at (0.5, -1) {};
		\node [style=none] (6) at (0.75, -1) {};
		\node [style=none] (7) at (1, -1) {};
		\node [style=none] (8) at (-0.5, -1) {};
		\node [style=none] (9) at (-0.75, -1) {};
		\node [style=none] (10) at (-1, -1) {};
		\node [style=none] (11) at (-1, 1) {};
		\node [style=none] (12) at (-0.75, 1) {};
		\node [style=none] (13) at (-0.5, 1) {};
		\node [style=Had] (14) at (0, -0) {$F^{\dagger}$};
	\end{pgfonlayer}
	\begin{pgfonlayer}{edgelayer}
		\draw (13.center) to (11.center);
		\draw (8.center) to (10.center);
		\draw (10.center) to (11.center);
		\draw (12.center) to (9.center);
		\draw (5.center) to (7.center);
		\draw (6.center) to (3.center);
		\draw (4.center) to (2.center);
		\draw (2.center) to (7.center);
		\draw (0.center) to (14);
		\draw (14) to (1.center);
	\end{pgfonlayer}
\end{tikzpicture} =$Fourier^{\dagger}$.

	\begin{tikzpicture}[scale=0.4]
	\begin{pgfonlayer}{nodelayer}
		\node [style=rn] (0) at (0, 0.5) {};
		\node [style=none] (1) at (0, -0.5) {};
		\node [style=none] (2) at (0.75, 1) {};
		\node [style=none] (3) at (0.75, -1) {};
		\node [style=none] (4) at (1, -1) {};
		\node [style=none] (5) at (1, 1) {};
		\node [style=none] (6) at (0.5, 1) {};
		\node [style=none] (7) at (0.5, -1) {};
		\node [style=none] (8) at (-0.5, -1) {};
		\node [style=none] (9) at (-0.75, -1) {};
		\node [style=none] (10) at (-1, -1) {};
		\node [style=none] (11) at (-1, 1) {};
		\node [style=none] (12) at (-0.75, 1) {};
		\node [style=none] (13) at (-0.5, 1) {};
	\end{pgfonlayer}
	\begin{pgfonlayer}{edgelayer}
		\draw (0) to (1.center);
		\draw (5.center) to (4.center);
		\draw (3.center) to (2.center);
		\draw (5.center) to (6.center);
		\draw (4.center) to (7.center);
		\draw (8.center) to (10.center);
		\draw (9.center) to (12.center);
		\draw (11.center) to (10.center);
		\draw (11.center) to (13.center);
	\end{pgfonlayer}
\end{tikzpicture} = $\ket{0}$ := $\sqrt{D} \left( \begin{array}{c} 1 \\ 0 \\ ... \\ 0
\end{array} \right)$  \hspace{1pt} ; \hspace{5pt} 
\begin{tikzpicture}[scale=0.4]
	\begin{pgfonlayer}{nodelayer}
		\node [style=gn] (0) at (0, 0.5) {};
		\node [style=none] (1) at (0, -0.5) {};
		\node [style=none] (2) at (0.75, 1) {};
		\node [style=none] (3) at (0.75, -1) {};
		\node [style=none] (4) at (1, -1) {};
		\node [style=none] (5) at (1, 1) {};
		\node [style=none] (6) at (0.5, 1) {};
		\node [style=none] (7) at (0.5, -1) {};
		\node [style=none] (8) at (-0.5, -1) {};
		\node [style=none] (9) at (-0.75, -1) {};
		\node [style=none] (10) at (-1, -1) {};
		\node [style=none] (11) at (-1, 1) {};
		\node [style=none] (12) at (-0.75, 1) {};
		\node [style=none] (13) at (-0.5, 1) {};
	\end{pgfonlayer}
	\begin{pgfonlayer}{edgelayer}
		\draw (0) to (1.center);
		\draw (5.center) to (4.center);
		\draw (3.center) to (2.center);
		\draw (5.center) to (6.center);
		\draw (4.center) to (7.center);
		\draw (8.center) to (10.center);
		\draw (9.center) to (12.center);
		\draw (11.center) to (10.center);
		\draw (11.center) to (13.center);
	\end{pgfonlayer}
\end{tikzpicture} = $\ket{+}$ := $ \left( \begin{array}{c} 1 \\ 1 \\ ... \\ 1
\end{array} \right)$ \hspace{1pt} ; \hspace{5pt} 
\begin{tikzpicture}[scale=0.4]
	\begin{pgfonlayer}{nodelayer}
		\node [style=rn] (0) at (0, -0.5) {};
		\node [style=none] (1) at (0, 0.5) {};
		\node [style=none] (2) at (0.75, 1) {};
		\node [style=none] (3) at (0.75, -1) {};
		\node [style=none] (4) at (1, -1) {};
		\node [style=none] (5) at (1, 1) {};
		\node [style=none] (6) at (0.5, 1) {};
		\node [style=none] (7) at (0.5, -1) {};
		\node [style=none] (8) at (-0.5, -1) {};
		\node [style=none] (9) at (-0.75, -1) {};
		\node [style=none] (10) at (-1, -1) {};
		\node [style=none] (11) at (-1, 1) {};
		\node [style=none] (12) at (-0.75, 1) {};
		\node [style=none] (13) at (-0.5, 1) {};
	\end{pgfonlayer}
	\begin{pgfonlayer}{edgelayer}
		\draw (0) to (1.center);
		\draw (5.center) to (4.center);
		\draw (3.center) to (2.center);
		\draw (5.center) to (6.center);
		\draw (4.center) to (7.center);
		\draw (8.center) to (10.center);
		\draw (9.center) to (12.center);
		\draw (11.center) to (10.center);
		\draw (11.center) to (13.center);
	\end{pgfonlayer}
\end{tikzpicture} = $\epsilon_{X}$:= $\bra{0}$  \hspace{1pt} ; \hspace{5pt} 
\begin{tikzpicture}[scale=0.4]
	\begin{pgfonlayer}{nodelayer}
		\node [style=gn] (0) at (0, -0.5) {};
		\node [style=none] (1) at (0, 0.5) {};
		\node [style=none] (2) at (0.75, 1) {};
		\node [style=none] (3) at (0.75, -1) {};
		\node [style=none] (4) at (1, -1) {};
		\node [style=none] (5) at (1, 1) {};
		\node [style=none] (6) at (0.5, 1) {};
		\node [style=none] (7) at (0.5, -1) {};
		\node [style=none] (8) at (-0.5, -1) {};
		\node [style=none] (9) at (-0.75, -1) {};
		\node [style=none] (10) at (-1, -1) {};
		\node [style=none] (11) at (-1, 1) {};
		\node [style=none] (12) at (-0.75, 1) {};
		\node [style=none] (13) at (-0.5, 1) {};
	\end{pgfonlayer}
	\begin{pgfonlayer}{edgelayer}
		\draw (0) to (1.center);
		\draw (5.center) to (4.center);
		\draw (3.center) to (2.center);
		\draw (5.center) to (6.center);
		\draw (4.center) to (7.center);
		\draw (8.center) to (10.center);
		\draw (9.center) to (12.center);
		\draw (11.center) to (10.center);
		\draw (11.center) to (13.center);
	\end{pgfonlayer}
\end{tikzpicture}  = $\epsilon_{Z}$:= $\bra{+}$ 

\begin{tikzpicture}[scale=0.4]
	\begin{pgfonlayer}{nodelayer}
		\node [style=rn] (0) at (0, -0) {};
		\node [style=none] (1) at (0, 1) {};
		\node [style=none] (2) at (1.5, 1) {};
		\node [style=none] (3) at (1.5, -1) {};
		\node [style=none] (4) at (1.75, -1) {};
		\node [style=none] (5) at (1.75, 1) {};
		\node [style=none] (6) at (1.25, 1) {};
		\node [style=none] (7) at (1.25, -1) {};
		\node [style=none] (8) at (-1.25, -1) {};
		\node [style=none] (9) at (-1.5, -1) {};
		\node [style=none] (10) at (-1.75, -1) {};
		\node [style=none] (11) at (-1.75, 1) {};
		\node [style=none] (12) at (-1.5, 1) {};
		\node [style=none] (13) at (-1.25, 1) {};
		\node [style=none] (14) at (-0.5, -1) {};
		\node [style=none] (15) at (0.5, -1) {};
	\end{pgfonlayer}
	\begin{pgfonlayer}{edgelayer}
		\draw (0) to (1.center);
		\draw (5.center) to (4.center);
		\draw (3.center) to (2.center);
		\draw (5.center) to (6.center);
		\draw (4.center) to (7.center);
		\draw (8.center) to (10.center);
		\draw (9.center) to (12.center);
		\draw (11.center) to (10.center);
		\draw (11.center) to (13.center);
		\draw (0) to (14.center);
		\draw (0) to (15.center);
	\end{pgfonlayer}
\end{tikzpicture} = $\delta_X :=\left( \begin{array}{c}
\mathbb{I}_{D\times D}\\
P_1(\mathbb{I}_{D\times D})\\
...\\
P_{D-1}(\mathbb{I}_{D\times D})  \end{array} \right)$ \hspace{2pt} ; \hspace{5pt}
\begin{tikzpicture}[scale=0.4]
	\begin{pgfonlayer}{nodelayer}
		\node [style=gn] (0) at (0, -0) {};
		\node [style=none] (1) at (0, 1) {};
		\node [style=none] (2) at (1.5, 1) {};
		\node [style=none] (3) at (1.5, -1) {};
		\node [style=none] (4) at (1.75, -1) {};
		\node [style=none] (5) at (1.75, 1) {};
		\node [style=none] (6) at (1.25, 1) {};
		\node [style=none] (7) at (1.25, -1) {};
		\node [style=none] (8) at (-1.25, -1) {};
		\node [style=none] (9) at (-1.5, -1) {};
		\node [style=none] (10) at (-1.75, -1) {};
		\node [style=none] (11) at (-1.75, 1) {};
		\node [style=none] (12) at (-1.5, 1) {};
		\node [style=none] (13) at (-1.25, 1) {};
		\node [style=none] (14) at (-0.5, -1) {};
		\node [style=none] (15) at (0.5, -1) {};
	\end{pgfonlayer}
	\begin{pgfonlayer}{edgelayer}
		\draw (0) to (1.center);
		\draw (5.center) to (4.center);
		\draw (3.center) to (2.center);
		\draw (5.center) to (6.center);
		\draw (4.center) to (7.center);
		\draw (8.center) to (10.center);
		\draw (9.center) to (12.center);
		\draw (11.center) to (10.center);
		\draw (11.center) to (13.center);
		\draw (0) to (14.center);
		\draw (0) to (15.center);
	\end{pgfonlayer}
\end{tikzpicture} = $\delta_Z :=\left( \begin{array}{cccc}
e_1 |& e_2| & ... &| e_D   \end{array} \right)$ \\

\begin{tikzpicture}[scale=0.4]
	\begin{pgfonlayer}{nodelayer}
		\node [style=none] (0) at (-0.5, -1) {};
		\node [style=none] (1) at (-0.5, 1) {};
		\node [style=rn] (2) at (-0.5, -0) {};
		\node [style=none] (3) at (-1.5, 1.5) {};
		\node [style=none] (4) at (-1.25, 1.5) {};
		\node [style=none] (5) at (-1, 1.5) {};
		\node [style=none] (6) at (4.75, 1.5) {};
		\node [style=none] (7) at (5, 1.5) {};
		\node [style=none] (8) at (5.25, 1.5) {};
		\node [style=none] (9) at (5, -1.5) {};
		\node [style=none] (10) at (4.75, -1.5) {};
		\node [style=none] (11) at (5.25, -1.5) {};
		\node [style=none] (12) at (-1.25, -1.5) {};
		\node [style=none] (13) at (-1.5, -1.5) {};
		\node [style=none] (14) at (-1, -1.5) {};
		\node [style=none] (15) at (2.5, -0) {$\alpha_1, ..., \alpha_{D-1}$};
	\end{pgfonlayer}
	\begin{pgfonlayer}{edgelayer}
		\draw (1.center) to (2);
		\draw (2) to (0.center);
		\draw (3.center) to (13.center);
		\draw (4.center) to (12.center);
		\draw (13.center) to (14.center);
		\draw (5.center) to (3.center);
		\draw (8.center) to (6.center);
		\draw (7.center) to (9.center);
		\draw (10.center) to (11.center);
		\draw (11.center) to (8.center);
	\end{pgfonlayer}
\end{tikzpicture} = $\Lambda_X(\alpha_1, \alpha_2, ..., \alpha_{D-1}):=\frac{1}{D}\left( \begin{array}{cccccc}
c_0 & c_{D-1} & c_{D-2} & ... & c_2 & c_1  \\
c_1 & c_0 & c_{D-1} & ... & c_3 & c_2 \\
c_2 & c_1 & c_0 & ... & c_4 & c_3 \\
... & ... & ... & ... & ... & ... \\
c_{D-1} & c_{D-2} & c_{D-3} & ... & c_1 & c_0 \end{array} \right)$ 

\begin{tikzpicture}[scale=0.4]
	\begin{pgfonlayer}{nodelayer}
		\node [style=none] (0) at (-0.5, -1) {};
		\node [style=none] (1) at (-0.5, 1) {};
		\node [style=gn] (2) at (-0.5, -0) {};
		\node [style=none] (3) at (-1.5, 1.5) {};
		\node [style=none] (4) at (-1.25, 1.5) {};
		\node [style=none] (5) at (-1, 1.5) {};
		\node [style=none] (6) at (4.75, 1.5) {};
		\node [style=none] (7) at (5, 1.5) {};
		\node [style=none] (8) at (5.25, 1.5) {};
		\node [style=none] (9) at (5, -1.5) {};
		\node [style=none] (10) at (4.75, -1.5) {};
		\node [style=none] (11) at (5.25, -1.5) {};
		\node [style=none] (12) at (-1.25, -1.5) {};
		\node [style=none] (13) at (-1.5, -1.5) {};
		\node [style=none] (14) at (-1, -1.5) {};
		\node [style=none] (15) at (2.5, -0) {$\alpha_1, ..., \alpha_{D-1}$};
	\end{pgfonlayer}
	\begin{pgfonlayer}{edgelayer}
		\draw (1.center) to (2);
		\draw (2) to (0.center);
		\draw (3.center) to (13.center);
		\draw (4.center) to (12.center);
		\draw (13.center) to (14.center);
		\draw (5.center) to (3.center);
		\draw (8.center) to (6.center);
		\draw (7.center) to (9.center);
		\draw (10.center) to (11.center);
		\draw (11.center) to (8.center);
	\end{pgfonlayer}
\end{tikzpicture} = $\Lambda_Z(\alpha_1, \alpha_2, ..., \alpha_{D-1}):=\left( \begin{array}{ccccc}
1 & 0 & 0 & ... & 0  \\
0 & e^{i \alpha_1} & 0 & ... & 0 \\
0 & 0 & e^{i \alpha_1} & ... & ... \\
... & ... & ... & ... & 0 \\
0 & 0 & ... & 0 & e^{i \alpha_{D-1}} \end{array} \right)$
 
 \begin{tikzpicture}[scale=0.4]
	\begin{pgfonlayer}{nodelayer}
		\node [style=rn] (0) at (0.4999999, 0.25) {};
		\node [style=none] (1) at (0.5, -1.5) {};
		\node [style=none] (2) at (-0.5, -1.5) {};
		\node [style=none] (3) at (-0.5, 1.25) {};
		\node [style=gn] (4) at (-0.4999999, -0.4999999) {};
		\node [style=none] (5) at (0.5, 1.25) {};
		\node [style=none] (6) at (-1.5, 1.5) {};
		\node [style=none] (7) at (-1.25, 1.5) {};
		\node [style=none] (8) at (-1, 1.5) {};
		\node [style=none] (9) at (1, 1.5) {};
		\node [style=none] (10) at (1.25, 1.5) {};
		\node [style=none] (11) at (1.5, 1.5) {};
		\node [style=none] (12) at (1.25, -1.75) {};
		\node [style=none] (13) at (1, -1.75) {};
		\node [style=none] (14) at (1.5, -1.75) {};
		\node [style=none] (15) at (-1.25, -1.75) {};
		\node [style=none] (16) at (-1.5, -1.75) {};
		\node [style=none] (17) at (-1, -1.75) {};
	\end{pgfonlayer}
	\begin{pgfonlayer}{edgelayer}
		\draw (3.center) to (4);
		\draw (4) to (0);
		\draw (0) to (5.center);
		\draw (0) to (1.center);
		\draw (4) to (2.center);
		\draw (6.center) to (16.center);
		\draw (7.center) to (15.center);
		\draw (16.center) to (17.center);
		\draw (8.center) to (6.center);
		\draw (11.center) to (9.center);
		\draw (10.center) to (12.center);
		\draw (13.center) to (14.center);
		\draw (14.center) to (11.center);
	\end{pgfonlayer}
\end{tikzpicture}= $CNOT_{a,b}:=\sum_{j,k=0}^{D-1}  \ket{k}\bra{j}_a \otimes \ket{k}\bra{k+j}_b =\left( \begin{array}{ccccc}
\mathbb{I}_{D \times D} & 0 & 0 & ... & 0  \\
0 & P_1(\mathbb{I}_{D \times D}) & 0 &   ... & 0  \\
0 & 0 & P_2(\mathbb{I}_{D \times D}) &   ... & ...  \\
... & ... & ... & ... & 0 \\
0 & 0  & ...  & 0 & P_{D-1}(\mathbb{I}_{D \times D})   \\
\end{array} \right)$
\caption{Hilbert space interpretation of the qudit ZX calculus elements.}
		\label{qudint}
\end{figure}
where  $P_j(\mathbb{I}_{D\times D})$ (j=1,2, ... D-1) are $D \times D$ matrices corresponding to the identity matrix $\mathbb{I}_{D\times D}$, with all its rows permuted to the right by j and where $e_i$ are the vectors which have one 1 in row $D\times (i-1)+i$ and D-1 zeros in all the other rows.

The $c_j$ elements of the $\Lambda_X$ matrix are defined by: $\ket{+}+\sum_{k=1}^{D-1}e^{i \alpha_k}\ket{+_k}=\frac{1}{\sqrt{D}}(c_0\ket{0}+\sum_{j=1}^{D-1} c_j \ket{j})$ where $\ket{+_k}$ are the D eigenvectors of the $X=\sum_{j=0}^{D-1}  \ket{j}\bra{j+1}$ matrix. 

We now proceed to show the universality of the qudit ZX calculus for quantum mechanics.
We will combine the following important results, due to Muthukrishnan and Stroud \cite{Mut00} and Brylinski \cite{Bry01} into a theorem: \\
\underline{Theorem 1:} The following set of qudit quantum gates are universal for quantum computing: \\
(a) The two following families of D-dimensional transforms (which are universal for single qudit quantum mechanics)\cite{Mut00}:
\begin{equation}
Z_j(b_0, b_1 ..., b_{D-1}):b_0 \ket{0}+b_1 \ket{1}+ ... +b_{D-1}  \ket{D-1} \mapsto \ket{j} 
\end{equation} for $j\in \{0, 1, ..., D-1 \}$.

\begin{equation}
X_j(\phi): b_0 \ket{0}+b_1 \ket{1}+ ... +b_{D-1}  \ket{D-1} \mapsto b_0 \ket{0}+ ... + e^{i \phi} b_j \ket{j} +b_{D-1}  \ket{D-1} 
\end{equation} for $j\in \{0, 1, ..., D-1 \}$.

(b) Any imprimitive 2 qudit gate\cite{Bry01}, where a 2 qudit gate is called imprimitive if there exist no single qudit gates S and T such that: $V=S \otimes T$ or $V=(S \otimes T)SWAP$ where $SWAP \ket{xy}=\ket{yx}$. \\

\underline{Theorem 2:} The qudit ZX calculus is universal for quantum mechanics. 
\begin{proof}
(a) Single qudit universality:

We can show that all 2D maps $Z_0$, $Z_1$, ..., $Z_{D-1}$, $X_0(\phi_0)$, $X_1(\phi_1)$, ..., and $X_{D-1}(\phi_{D-1})$ are included in the qudit ZX calculus.

One can easily check that (up to global phase): $X_0(\phi_0)=\Lambda_Z(-\phi_0, ..., -\phi_0)$, $X_j(\phi_j)=\Lambda_Z(0, ..., 0 ,\alpha_j=\phi_j, 0, ..., 0)$ for $j \neq 0$.

Next, we show that the d maps $Z_j(b_0, b_1, ... ,b_{D-1})$ can be written (up to global phase) as $\Lambda_X(\alpha_1,\alpha_2, ..., \alpha_{D-1})$ for some set $\{ \alpha_1, \alpha_2, ..., \alpha_D \}$.

First of all, note that since $\Lambda_X(\alpha_1,\alpha_2, ..., \alpha_{D-1})$ is unitary for any values of $\{ \alpha_1,\alpha_2, ..., \alpha_{D-1} \}$:

\begin{equation}
det(\Lambda_X(\alpha_1,\alpha_2, ..., \alpha_{D-1}))=1 \neq 0
\end{equation}

This means that there exists a unique solution set $\{ b_0,b_1, ..., b_{D-1} \}$ to the equation:

\begin{equation}\label{lambdaa}
\Lambda_X(\alpha_1,\alpha_2, ..., \alpha_{D-1}) . (b_0, b_1, ... ,b_{D-1})^{T} = e_j
\end{equation}

for each vector $e_j =(0, ...,0, 1, 0, ..., 0)^{T}$ with a single 1 in the $j^{th}$ column.

Using the definition of $\Lambda_X$, this means that there exists a unique set of $\{ b_0,b_1, ..., b_{D-1} \}$ such that:
\begin{align}\label{uniqs1}
&c_j b_0+c_{j-1} b_1+ ...+ c_0 b_j+c_{D-1} b_{j+1}+ ... +c_{j+1} b_{D-1}= D \\
&c_k b_0+c_{k-1} b_1+ ...+ c_0 b_k+c_{D-1} b_{k+1}+ ... +c_{k+1} b_{D-1}=0; \forall k \neq j \label{uniqs2}
\end{align}

Since we know that: $ \sum_{k=0}^{D-1} \eta^k=0$,
where $\eta= e^{\frac{2 \pi i}{D}}$, and that $\sum_{k=0}^{D-1} b_k^{\star} b_k =1$ this means that there is a unique solution to the set of equations:
\begin{equation}\label{ccs}
 c_j= D b_0^{\star},  c_{j-1}= D b_1^{\star}, ...,  c_0=D b_j^{\star}, ...,  c_{j+1}=D b_{D-1}^{\star} 
\end{equation}
consistent with (\ref{uniqs1},\ref{uniqs2}). 

But all the $b_k^{\star}$ ($k=0,1, ..., D-1$) are complex numbers of unit norm and by definition the $c_k$ can be written as: $c_k=1+\sum_{l=1}^{D-1} \eta^{r_k(l)} e^{i \alpha_l} $ where $r_k$ permutes the entries l (there is one $r_k$ for each k). Therefore, up to global phase, it is always possible to find values for $\{\alpha_1, ..., \alpha_{D-1} \}$ such that (\ref{ccs}) is satisfied. This means that each map $Z_D(b_0, ..., b_{D-1})$ (each one corresponding to a value of j in (\ref{lambdaa})) can be written in the form $\Lambda_X(\alpha_1,\alpha_2, ..., \alpha_{D-1})$ for some set $\{ \alpha_1, \alpha_2, ..., \alpha_D \}$.     

Using theorem 1, this shows that the qudit ZX calculus contains all single qudit unitary transformations. \\

(b) The qudit CNOT gate is an imprimitive 2 qudit gate which is contained within the qudit calculus. \\

 Note also that any map from n qudits to m qudits can be constructed by using diagrammatic map-state duality \cite{Ab04}.
Therefore, any qudit quantum state and (post-selected) measurement and any quantum gate can be written in the qudit ZX calculus and therefore it is \textbf{universal} for quantum mechanics. 
\end{proof}

To illustrate the proof for single qudit universality, note that in the qutrit case, we can explicitly find an assignment of the $\alpha$ values such that (up to global phase):

\small
\begin{align}
Z_0(b_0, b_1,b_2)=\Lambda_X(-i \log & \left( \frac{(b_0+b_1+b_2) (b_0 \eta-b_1 (\eta+1)+b_2)}{\eta \left(b_0^2 (-\eta) (\eta+1)-b_0
   (b_1+b_2)+b_1^2+b_1 b_2 \eta (\eta+1)+b_2^2\right)}\right), \nonumber \\
     -i &\log \left(\frac{(b_0+b_1+b_2) (b_0 \eta+b_1-b_2 (\eta+1))}{\eta \left(b_0^2 (-\eta) (\eta+1)-b_0
   (b_1+b_2)+b_1^2+b_1 b_2 \eta (\eta+1)+b_2^2\right)}\right))
\end{align}
\begin{align}
Z_1(b_0, b_1,b_2)=\Lambda_X(-i \log & \left( \frac{(b_0+b_1+b_2) (b_0+b_1 \eta-b_2 (\eta+1))}{\eta \left(b_0^2+b_0 (b_2 \eta (\eta+1)-b_1)-(b_1
   \eta+b_1-b_2) (b_1 \eta+b_2)\right)}\right), \nonumber\\
     -i &\log \left(-\frac{(b_0+b_1+b_2) (b_0 \eta+b_0-b_1 \eta-b_2)}{\eta \left(b_0^2+b_0 (b_2 \eta
   (\eta+1)-b_1)-(b_1 \eta+b_1-b_2) (b_1 \eta+b_2)\right)}\right))
\end{align}
\begin{align}
Z_2(b_0, b_1,b_2)=\Lambda_X(-i \log & \left( -\frac{(b_0+b_1+b_2) (b_0 \eta+b_0-b_1-b_2 \eta)}{\eta \left(b_0^2+b_0 (b_1 \eta
   (\eta+1)-b_2)+(b_1+b_2 \eta) (b_1-b_2 (\eta+1))\right)}\right), \nonumber \\ 
    -i &\log \left(\frac{(b_0+b_1+b_2) (b_0-b_1 (\eta+1)+b_2 \eta)}{\eta \left(b_0^2+b_0 (b_1 \eta
   (\eta+1)-b_2)+(b_1+b_2 \eta) (b_1-b_2 (\eta+1))\right)}\right))
\end{align}
\normalsize
where $\eta= e^{\frac{2 \pi i}{3}}$.

By construction, any equation which can be shown to be true using the qudit ZX calculus is true in quantum mechanics so the qudit ZX calculus is \textbf{sound} for quantum mechanics. Moreover, extending the qudit ZX calculus to account for mixed states and general quantum evolution described by completely positive maps can be achieved by using the same standard constructions \cite{Sel07, Per10, Dun10} as in the qubit case. 

We know that \cite{Back12} the qubit ZX calculus is \textbf{complete} for qubit stabilizer quantum theory, in the sense that any two equivalent qubit stabilizer processes can be shown to be equal by using the qubit ZX calculus. Backens' proof of this result \cite{Back12}, however, relies on results for qubit graph states and it is unclear whether it can be generalized to show completeness of the qudit ZX calculus for qudit stabilizer theory. Therefore, we leave this as an open question: \\
\textit{Is the qudit ZX calculus, with additional rules analogous to the Euler decomposition of the Hadamard vertex, complete for qudit stabilizer quantum mechanics? If it is not, then which other rules need to be added for completeness?} \\  

Another important question is how the qudit and qubit ZX calculi are related. More generally, it would be interesting to understand exactly how the ZX calculus for qudits of dimension m is related to the ZX calculus of dimension $n>m$. Perhaps, we could introduce maps which $``create"$ and $``annihilate"$ dimensions. This could lead to an interesting structure and provide insight into the relationship between qubit and qudit quantum mechanics.

\section{Mutually unbiased qudit theories}

One of the main goals of the present article is to use the abstract structures we introduced to study the foundation of quantum theory. In this respect, we aim to define a class of theories which exhibit many key features of quantum mechanics, within a single mathematical framework. 

Therefore, we will generalize the previous approach of studying mutually unbiased qubit theories using dagger compact symmetric monoidal categories \cite{Ed11, Edw10} to the case of qudits.

\underline{Definition:} A \textbf{mutually unbiased qudit theory} is a dagger symmetric monoidal category with observable structures such that:

(i) The objects of the category are the unit and finite tensor products of qudit-like systems Q.

(ii) The observables on a given object are all mutually unbiased, have the same number of eigenstates and have the same phase groups. 

(iii) All states of Q are eigenstates of some observable.

We will study mutually unbiased qudit theories for dimensions higher than two. In the following two sections, we analyze in detail two key examples of mutually unbiased qudit theories: \textit{qudit stabilizer quantum mechanics} and \textit{Spekkens-Screiber theory for dits}. 
  
\section{Picturing stabilizer quantum mechanics}

We define the process category \textbf{DStab} as the $\dagger$-compact symmetric monoidal subcategory of \textbf{FHilb} corresponding to qudit stabilizer quantum mechanics (see Appendix B) which is generated by the unit, n-fold tensor products of $\mathbb{C}^D$, single qudit Clifford operations and the quantum copying and deleting maps. \textbf{DStab} can be depicted using the qudit ZX diagrams, where the allowed phases are restricted according to the phase group.

In the case of the standard qubit stabilizer quantum mechanics, the phase group is the cyclic group $\mathbb{Z}_4$, which is a finite subgroup of the quantum qubit phase group $S^1$ (the circle group). Since the unbiased circles for the Z and X observables coincide on the points corresponding to $\ket{+i}$ and $\ket{-i}$, we can completely picture single qubit stabilizer quantum theory using the Bloch sphere.

Can one find an analogous picture for qutrit quantum mechanics? \\
Let $\{ \ket{0}$, $\ket{1}$, $\ket{2} \}$ and $\{ \ket{+}$, $\ket{\top}$, $\ket{\perp} \}$ be the eigenbases for the qutrit Z and X observables respectively. Then the unbiased states for the Z and X observable:
\begin{equation}
\ket{\{\alpha_1,\alpha_2 \}_Z}= \ket{0}+ e^{i \alpha} \ket{1}+e^{i \alpha_2} \ket{2} \hspace{5pt} ; \hspace{5pt} \ket{\{\alpha_1,\alpha_2 \}_X}= \ket{+}+ e^{i \alpha} \ket{\top}+e^{i \alpha_2} \ket{\perp}
\end{equation}
under pairwise addition of phases form a torus group $S^1 \times S^1$. 

All the single qutrit stabilizer states, corresponding to the eigenstates of the qutrit X, Z, XZ and $XZ^2$ operators, can be written as unbiased states for either the Z basis or the X basis since:
\begin{equation}
\begin{split}
&\ket{0}=\ket{\{0, 0 \}_X} , \ket{1}= \ket{\{ \frac{4 \pi}{3},\frac{2 \pi}{3} \}_X}, \ket{2}= \ket{\{ \frac{2 \pi}{3},\frac{4 \pi}{3} \}_X}; \\
&\ket{+}=\ket{\{0, 0 \}_Z} , \ket{\top}= \ket{\{\frac{2 \pi}{3}, \frac{4 \pi}{3} \}_Z}, \ket{\perp}= \ket{\{\frac{4 \pi}{3}, \frac{2 \pi}{3} \}_Z}; \\
&\ket{-}=\ket{\{\frac{4 \pi}{3}, \frac{4 \pi}{3} \}_Z}=\ket{\{\frac{2 \pi}{3}, \frac{2 \pi}{3} \}_X} , \ket{\vdash}= \ket{\{0, \frac{2 \pi}{3} \}_Z}=\ket{\{ \frac{4 \pi}{3},0 \}_X}, \ket{\dashv}= \ket{\{\frac{2 \pi}{3}, 0 \}_Z}=\ket{\{0, \frac{4 \pi}{3} \}_X}; \\
& \ket{\times}=\ket{\{\frac{2 \pi}{3}, \frac{2 \pi}{3} \}_Z}=\ket{\{\frac{4 \pi}{3}, \frac{4 \pi}{3} \}_X} , \ket{\leftthreetimes}= \ket{\{ \frac{4 \pi}{3},0 \}_Z}=\ket{\{ \frac{2 \pi}{3},0 \}_X}, \ket{\rightthreetimes}= \ket{\{0,\frac{4 \pi}{3} \}_Z}=\ket{\{0,\frac{2 \pi}{3} \}_X}
\end{split}
\end{equation}
\begin{figure}[h]
	\centering
	\begin{subfigure}{.5\textwidth}
  \raggedleft 
  \includegraphics[ totalheight=0.3\textheight]{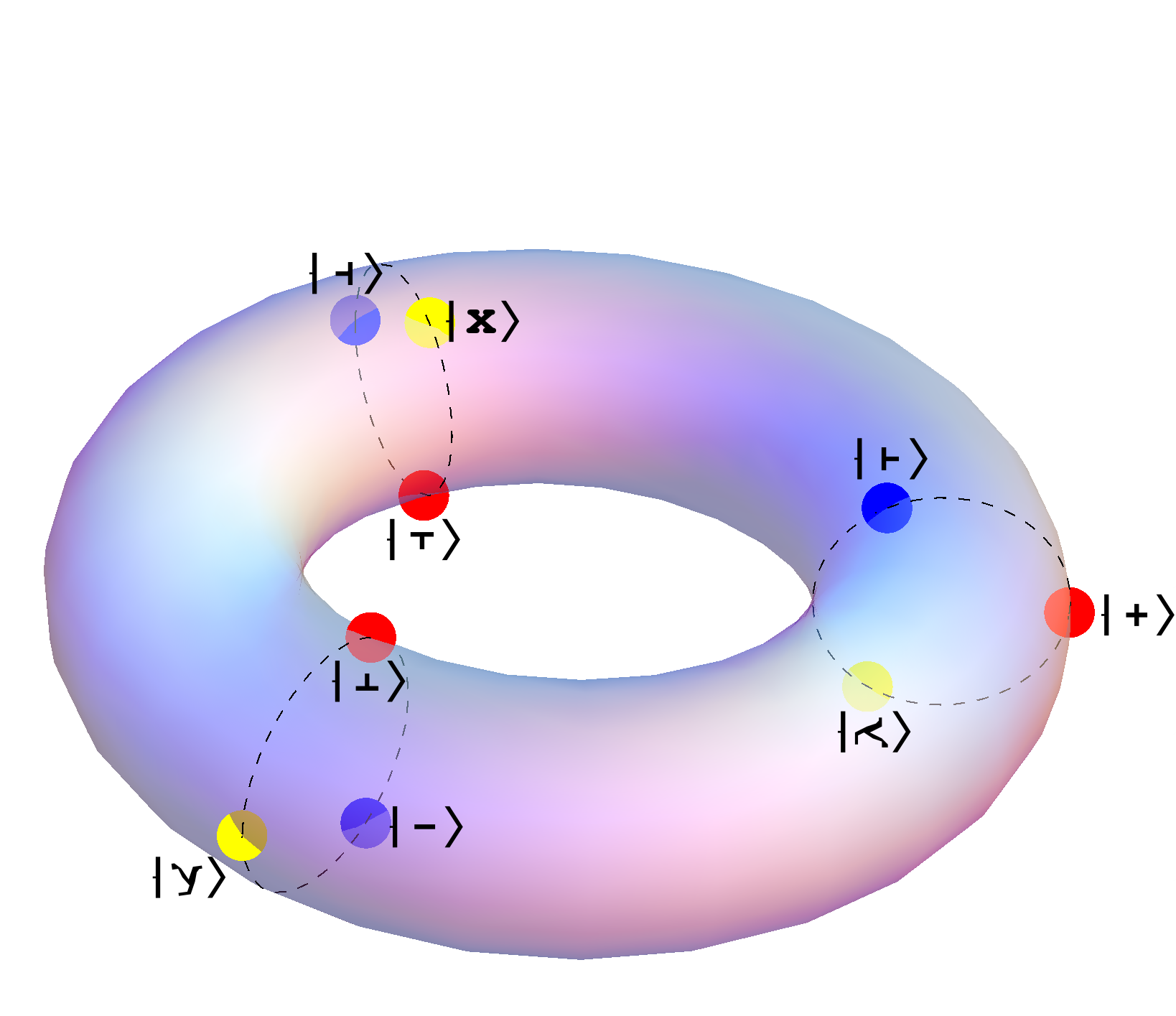}	\caption{Qutrit stabilizer states on the unbiased \\ torus for the Z observable.}
  \label{torusx}
\end{subfigure}%
\begin{subfigure}{.5\textwidth}
  \raggedright
  \includegraphics[ totalheight=0.25\textheight]{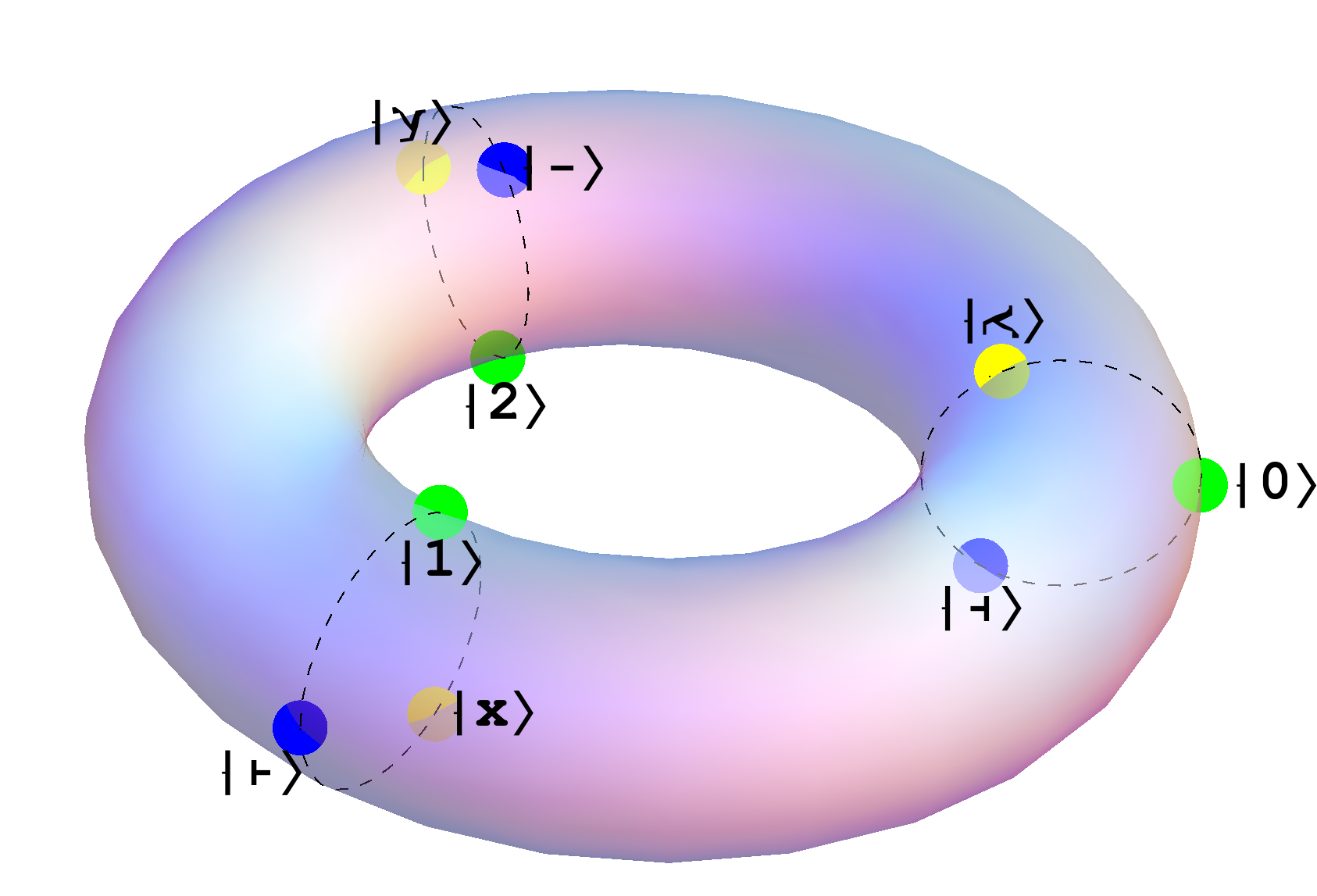}	\caption{Qutrit stabilizer states on the unbiased \\ torus for the X observable.}
		\label{torusz}
\end{subfigure}
\caption{Depicting qutrit stabilizer theory on two tori.}
\end{figure}

Single qutrit stabilizer operations take subsets of these 12 states to other subsets of these 12 states. This shows that the phase group for qutrit stabilizer quantum mechanics is \textbf{$\mathbb{Z}_3 \times \mathbb{Z}_3$}. \\

Therefore, single qutrit stabilizer quantum theory can be pictured using 12 points on two toruses, which is a direct generalization of the Bloch sphere case, where the 4 elements on each of the two unbiased circles (coinciding on two elements) visualized in three dimensions are replaced by \textbf{9 elements on each of two unbiased toruses coinciding on six points} (the blue and yellow points in Figures (\ref{torusx}, \ref{torusz})).

In fact, this picture can easily be generalized to higher dimensional qudit stabilizer theories for prime dimensions. In that case, the single qudit states of qudit stabilizer quantum theory correspond to the vectors in the D+1 mutually unbiased eigenbases of the single qudit operators: $X, Z, XZ, XZ^2, ...,XZ^{D-1}$. The mutually unbiased points with respect to each of these bases forms a D-torus. If we chose an observable structure, whose eigenstates are a basis, then all the other stabilizer states are on the unbiased D-torus of the chosen basis. In this way, qudit stabilizer theory for prime dimension D can be pictured using \textbf{$D^2$ points on each of two D-toruses} (unbiased toruses for the Z and X operators for example), which coincide on $D^2-D$ points and can be visualized in D+1 dimensions .

In general, the phase group for qudit stabilizer quantum theory of dimension $D>3$ is an Abelian subgroup of the group $\mathbb{Z}_D \times \mathbb{Z}_D \times ... \times \mathbb{Z}_D $ (D-1 times). In fact, every finite dimensional closed subgroup of the torus group is isomorphic to a product of finite cyclic groups. Therefore, the phase group for mutually unbiased qudit theories which are also subtheories of quantum mechanics must be of the form  $Z_{n_1} \times Z_{n_2} \times ... \times Z_{n_k}$ for positive integers $n_1, ..., n_k$. In further work, we will study how these integers $n_1, ..., n_k$ for stabilizer phase groups depend on the dimension D. In general, we would like a physical classification of all the mutually unbiased qudit subtheories of quantum mechanics in terms of $n_1, ..., n_k$. Once we have determined their \textbf{phase group}, the qudit ZX calculus allows us to fully describe these physical theories.

\section{Depicting Spekkens-Schreiber toy theory for dits}

We define the category \textbf{FRel} whose objects are finite sets and whose morphisms are relations. By taking the Cartesian product of sets as the tensor product, the single element set $\{ \star \}$ as the identity object and the relational converse as the dagger, \textbf{FRel} can be viewed as a SMC with dagger structure.  

We can then define the category \textbf{DSpek} as a subcategory of \textbf{FRel} whose objects are the single element set $I=\{ \star\}$ and n-fold Cartesian products of the $D^2$-element set: $\mathcal{D} :=\{1, 2, ..., D^{2} \}$. 

The morphisms of \textbf{DSpek} are those generated by composition, Cartesian product and relational converse from the following relations:

(a) All $(D^2)!$ permutations $\sigma_i:\mathcal{D} \rightarrow \mathcal{D}$ of the $D^2$- element set.

(b) The copying relation: $\delta_Z: \mathcal{D} \rightarrow \mathcal{D} \times \mathcal{D}$ defined as:

\begin{center}
\resizebox{0.85\textwidth}{0.13\textheight}{
\begin{tabular}{|l|l|l|l|l|l|l|l|l|l|l|l|l|l|l|l|}
\hline
1   & 2   & ... & D   &     &     &     &      &     &     &     &     &          &          &     &         \\ \hline
D   & 1   & ... & D-1 &     &     &     &      &     &     &     &     &          &          &     &         \\ \hline
... & ... & ... & ... &     &     &     &      &     &     &     &     &          &          &     &         \\ \hline
2   & 3   & ... & 1   &     &     &     &      &     &     &     &     &          &          &     &         \\ \hline
    &     &     &     & D+1 & D+2 & ... & 2D   &     &     &     &     &          &          &     &         \\ \hline
    &     &     &     & 2D  & D+1 & ... & 2D-1 &     &     &     &     &          &          &     &         \\ \hline
    &     &     &     & ... & ... & ... & ...  &     &     &     &     &          &          &     &         \\ \hline
    &     &     &     & D+2 & D+3 & ... & D+1  &     &     &     &     &          &          &     &         \\ \hline
    &     &     &     &     &     &     &      & ... & ... & ... & ... &          &          &     &         \\ \hline
    &     &     &     &     &     &     &      & ... & ... & ... & ... &          &          &     &         \\ \hline
    &     &     &     &     &     &     &      & ... & ... & ... & ... &          &          &     &         \\ \hline
    &     &     &     &     &     &     &      & ... & ... & ... & ... &          &          &     &         \\ \hline
    &     &     &     &     &     &     &      &     &     &     &     & D(D-1)+1 & D(D-1)+2 & ... & $D^2$   \\ \hline
    &     &     &     &     &     &     &      &     &     &     &     & $D^2$    & D(D-1)+1 & ... & $D^2$-1 \\ \hline
    &     &     &     &     &     &     &      &     &     &     &     & ...      & ...      & ... & ...     \\ \hline
    &     &     &     &     &     &     &      &     &     &     &     & D(D-1)+2 & D(D-1)+3 & ... &  D(D-1)+1      \\ \hline
\end{tabular}}
\end{center}

where there is x in the (y,z) location of the grid iff $\delta_Z: x \sim (y,z)$.

(c) The deleting relation: $\epsilon_Z:\mathcal{D} \rightarrow I $ defined as: $\{ 1,D+1, 2D +1, ..., D(D-1)+1\} \sim \star$.

(d) The relevant unit, associativity and symmetry natural isomorphisms.

If we interpret relations from I to n-fold tensor products of $\mathcal{D}$ as epistemic states on phase space then this category corresponds to Spekkens-Schreiber theory for dits with only states of maximal knowledge. Adding the maximally mixed state $\perp_D :: \{ \star \} \sim \{1, 2, ..., D^2 \}$ to \textbf{DSpek} yields the category \textbf{MDSpek}, corresponding to Spekkens-Schreiber theory for dits of dimension D (See Appendix C).

\textbf{DSpek} and \textbf{MDSpek} inherit symmetric monoidal and $\dagger$-compact structure from \textbf{FRel} since we can define Bell states (corresponding to compact structures) as:

\begin{equation}
\mu_D := \delta_Z \circ \epsilon_Z^{\dagger}: I \rightarrow \mathcal{D} \times \mathcal{D} :: \star \sim \{(1,1),(2,2), ..., (D^2,D^2) \}
\end{equation} 

We can define the other copying map as:
\begin{equation}
\delta_X= (\Pi_{k=1}^{D-1} \sigma_{(k+1,(kD)+1)} \otimes \Pi_{k=1}^{D-1} \sigma_{(k+1,(kD)+1)}) \circ \delta_Z \circ (\Pi_{k=1}^{D-1} \sigma_{(k+1,(kD)+1)})
\end{equation}
where $\sigma_{(j,k)}$ permutes entries j and k of the input $D^2$-element set (epistemic state).
This map is explicitly: $\delta_X: \mathcal{D} \rightarrow \mathcal{D} \times \mathcal{D}$ such that: $\delta_X: x \sim (y,z)$ iff there is x in the (y,z) location of the following grid:

\begin{center}
\resizebox{0.85\textwidth}{0.13\textheight}{
\begin{tabular}{|l|l|l|l|l|l|l|l|l|l|l|l|l|l|l|l|}
\hline
1        &          &     &       & D+1  &      &     &     & ... & ... & ... & ... & D(D-1)+1 &          &     &        \\ \hline
         & 2        &     &       &      & D+2  &     &     & ... & ... & ... & ... &          & D(D-1)+2 &     &        \\ \hline
         &          & ... &       &      &      & ... &     & ... & ... & ... & ... &          &          & ... &        \\ \hline
         &          &     & D     &      &      &     & 2D  & ... & ... & ... & ... &          &          &     & $D^2$  \\ \hline
D(D-1)+1 &          &     &       & 1    &      &     &     & ... & ... & ... & ... & D(D-2)+1 &          &     &        \\ \hline
         & D(D-1)+2 &     &       &      & 2    &     &     & ... & ... & ... & ... &          & D(D-2)+2 &     &        \\ \hline
         &          & ... &       &      &      & ... &     & ... & ... & ... & ... &          &          & ... &        \\ \hline
         &          &     & $D^2$ &      &      &     & D   & ... & ... & ... & ... &          &          &     & D(D-1) \\ \hline
...      & ...      & ... & ...   & ...  & ...  & ... & ... & ... & ... & ... & ... & ...      & ...      & ... & ...    \\ \hline
...      & ...      & ... & ...   & ...  & ...  & ... & ... & ... & ... & ... & ... & ...      & ...      & ... & ...    \\ \hline
...      & ...      & ... & ...   & ...  & ...  & ... & ... & ... & ... & ... & ... & ...      & ...      & ... & ...    \\ \hline
...      & ...      & ... & ...   & ...  & ...  & ... & ... & ... & ... & ... & ... & ...      & ...      & ... & ...    \\ \hline
D+1      &          &     &       & 2D+1 &      &     &     & ... & ... & ... & ... & 1        &          &     &        \\ \hline
         & D+2      &     &       &      & 2D+2 &     &     & ... & ... & ... & ... &          & 2        &     &        \\ \hline
         &          & ... &       &      &      & ... &     & ... & ... & ... & ... &          &          & ... &        \\ \hline
         &          &     & 2D    &      &      &     & 3D  & ... & ... & ... & ... &          &          &     & D      \\ \hline
\end{tabular}}
\end{center}

and the other erasing map as: $\epsilon_X:\mathcal{D} \rightarrow I $ such that: $\{ 1, 2, 3, ..., D\} \sim \star$. It is easy to check that this then gives us two strongly complementary observable structures, analogous to the Z and X observable structures in quantum theory. 

In fact, we can use the fact that \textbf{3Spek} can be depicted in the qutrit ZX calculus to provide a novel proof of the following known result:

\underline{Theorem 3}\cite{Gro06,Scsp12}: Spekkens-Schreiber theory for trits is operationally equivalent to stabilizer theory for qutrits. 
\begin{proof} 
We can define the Z and X observable structures $(\mathcal{D},\delta_Z^{(trit)},\epsilon_Z^{(trit)})$ and $(\mathcal{D},\delta_X^{(trit)},\epsilon_X^{(trit)})$ for \textbf{3Spek} as described above. 

The Z observable structure has three classical states:

\begin{equation}
z_0 :: \star \sim \{1,2,3 \} \hspace{5pt}; \hspace{5pt} z_1 :: \star \sim \{4,5,6 \} \hspace{5pt}; \hspace{5pt} z_2 :: \star \sim \{7,8,9 \}
\end{equation}
and nine unbiased states:
\begin{equation*}
x_0 :: \star \sim \{1,4,7 \} \hspace{5pt} ; \hspace{5pt} x_1 :: \star \sim \{2,5,8 \} \hspace{5pt} ; \hspace{5pt} x_2 :: \star \sim \{3,6,9 \}
\end{equation*}
\begin{equation*}
(xz)_0 :: \star \sim \{1,6,8 \} \hspace{5pt} ; \hspace{5pt} (xz)_1 :: \star \sim \{2,4,9 \} \hspace{5pt} ; \hspace{5pt} (xz)_2 :: \star \sim \{3,5,7 \}
\end{equation*}
\begin{equation}(xz^2)_0 :: \star \sim \{1,5,9 \} \hspace{5pt} ; \hspace{5pt} (xz^2)_1 :: \star \sim \{2,6,7 \} \hspace{5pt} ; \hspace{5pt} (xz^2)_2 :: \star \sim \{3,4,6 \}
\end{equation}

The X observable also has 3 classical states $x_0, x_1, x_2$ and nine unbiased states $z_0, z_1, z_2,$ $(xz)_0,(xz)_1,(xz)_2,$ $(xz^2)_0, (xz^2)_1$ and $(xz^2)_2$.

Similarly, one could define two other observable structures corresponding to ``XZ" and ``X$Z^2$" which each have three classical states and nine unbiased states. Therefore, \textbf{3Spek} contains 4 observable structures and 12 single trit states. Also, there are 12 (measurement) effects, corresponding to taking the converse relations of the 12 states.

If we define phase maps: $\Lambda_Z(\psi):=(\delta_Z^{(trit)})^{\dagger}(\psi \otimes 1_{\mathcal{D}}):\mathcal{D}\rightarrow \mathcal{D}$ and $\Lambda_X(\psi):=(\delta_X^{(trit)})^{\dagger}(\psi \otimes 1_{\mathcal{D}}):\mathcal{D}\rightarrow \mathcal{D}$, then it is clear that the \textit{phase group} of \textbf{3Spek} is $\mathbb{Z}_3 \times \mathbb{Z}_3$. \\

Therefore, both \textbf{3Spek} and \textbf{3Stab} can be expressed in the qutrit ZX calculus as mutually unbiased qutrit theories with twelve states and phase group $\mathbb{Z}_3 \times \mathbb{Z}_3$. This uniquely determines all the allowable preperations, measurements and transformations (compositions of spiders with phases adding according to $\mathbb{Z}_3 \times \mathbb{Z}_3$).  

Explicitly, we can associate the twelve single trit states of \textbf{3Spek} with those of \textbf{3Stab} according to: 
\begin{equation}
\begin{split}
&[\![ z_0 ]\!]\equiv  \ket{0}=\begin{tikzpicture}[scale=0.65]
	\begin{pgfonlayer}{nodelayer}
		\node [style=rn] (0) at (0, 0.5) {};
		\node [style=none] (1) at (0, -0.5) {};
		\node [style=none] (2) at (0.8, 0.5) {0,0};
	\end{pgfonlayer}
	\begin{pgfonlayer}{edgelayer}
		\draw (0) to (1.center);
	\end{pgfonlayer}
\end{tikzpicture}; \hspace{5pt} [\![ z_1 ]\!]\equiv  \ket{1}= \begin{tikzpicture}[scale=0.65]
	\begin{pgfonlayer}{nodelayer}
		\node [style=rn] (0) at (0, 0.5) {};
		\node [style=none] (1) at (0, -0.5) {};
		\node [style=none] (2) at (1, 0.5) {$\frac{4 \pi}{3},\frac{2 \pi}{3}$};
	\end{pgfonlayer}
	\begin{pgfonlayer}{edgelayer}
		\draw (0) to (1.center);
	\end{pgfonlayer}
\end{tikzpicture}; \hspace{5pt} [\![ z_2 ]\!]\equiv  \ket{2}= \begin{tikzpicture}[scale=.65]
	\begin{pgfonlayer}{nodelayer}
		\node [style=rn] (0) at (0, 0.5) {};
		\node [style=none] (1) at (0, -0.5) {};
		\node [style=none] (2) at (1, 0.5) {$\frac{2 \pi}{3},\frac{4 \pi}{3}$};
	\end{pgfonlayer}
	\begin{pgfonlayer}{edgelayer}
		\draw (0) to (1.center);
	\end{pgfonlayer}
\end{tikzpicture}; \\
&[\![ x_0 ]\!]\equiv \ket{+}=\begin{tikzpicture}[scale=0.65]
	\begin{pgfonlayer}{nodelayer}
		\node [style=gn] (0) at (0, 0.5) {};
		\node [style=none] (1) at (0, -0.5) {};
		\node [style=none] (2) at (0.8, 0.5) {0,0};
	\end{pgfonlayer}
	\begin{pgfonlayer}{edgelayer}
		\draw (0) to (1.center);
	\end{pgfonlayer}
\end{tikzpicture}; \hspace{5pt} [\![ x_1 ]\!]\equiv \ket{\top}= \begin{tikzpicture}[scale=0.65]
	\begin{pgfonlayer}{nodelayer}
		\node [style=gn] (0) at (0, 0.5) {};
		\node [style=none] (1) at (0, -0.5) {};
		\node [style=none] (2) at (1, 0.5) {$\frac{2 \pi}{3},\frac{4 \pi}{3}$};
	\end{pgfonlayer}
	\begin{pgfonlayer}{edgelayer}
		\draw (0) to (1.center);
	\end{pgfonlayer}
\end{tikzpicture}; \hspace{5pt} [\![ x_2 ]\!]\equiv \ket{\perp}= \begin{tikzpicture}[scale=0.65]
	\begin{pgfonlayer}{nodelayer}
		\node [style=gn] (0) at (0, 0.5) {};
		\node [style=none] (1) at (0, -0.5) {};
		\node [style=none] (2) at (1, 0.5) {$\frac{4 \pi}{3},\frac{2 \pi}{3}$};
	\end{pgfonlayer}
	\begin{pgfonlayer}{edgelayer}
		\draw (0) to (1.center);
	\end{pgfonlayer}
\end{tikzpicture}; \\
&[\![ (xz)_0 ]\!]\equiv \ket{-}=\begin{tikzpicture}[scale=0.65]
	\begin{pgfonlayer}{nodelayer}
		\node [style=gn] (0) at (0, 0.5) {};
		\node [style=none] (1) at (0, -0.5) {};
		\node [style=none] (2) at (1, 0.5) {$\frac{4 \pi}{3},\frac{4 \pi}{3}$};
	\end{pgfonlayer}
	\begin{pgfonlayer}{edgelayer}
		\draw (0) to (1.center);
	\end{pgfonlayer}
\end{tikzpicture}; \hspace{5pt} [\![ (xz)_1 ]\!]\equiv \ket{\vdash}= \begin{tikzpicture}[scale=0.65]
	\begin{pgfonlayer}{nodelayer}
		\node [style=gn] (0) at (0, 0.5) {};
		\node [style=none] (1) at (0, -0.5) {};
		\node [style=none] (2) at (1, 0.5) {$0,\frac{2 \pi}{3}$};
	\end{pgfonlayer}
	\begin{pgfonlayer}{edgelayer}
		\draw (0) to (1.center);
	\end{pgfonlayer}
\end{tikzpicture}; \hspace{5pt} [\![ (xz)_2 ]\!]\equiv \ket{\dashv}= \begin{tikzpicture}[scale=0.65]
	\begin{pgfonlayer}{nodelayer}
		\node [style=gn] (0) at (0, 0.5) {};
		\node [style=none] (1) at (0, -0.5) {};
		\node [style=none] (2) at (1, 0.5) {$\frac{2 \pi}{3},0$};
	\end{pgfonlayer}
	\begin{pgfonlayer}{edgelayer}
		\draw (0) to (1.center);
	\end{pgfonlayer}
\end{tikzpicture}; \\
& [\![ (xz^2)_0 ]\!]\equiv \begin{tikzpicture}[scale=0.65]
	\begin{pgfonlayer}{nodelayer}
		\node [style=gn] (0) at (0, 0.5) {};
		\node [style=none] (1) at (0, -0.5) {};
		\node [style=none] (2) at (1, 0.5) {$\frac{2 \pi}{3},\frac{2 \pi}{3}$};
	\end{pgfonlayer}
	\begin{pgfonlayer}{edgelayer}
		\draw (0) to (1.center);
	\end{pgfonlayer}
\end{tikzpicture}; \hspace{5pt} [\![ (xz^2)_1 ]\!]\equiv  \ket{\leftthreetimes}= \begin{tikzpicture}[scale=0.65]
	\begin{pgfonlayer}{nodelayer}
		\node [style=gn] (0) at (0, 0.5) {};
		\node [style=none] (1) at (0, -0.5) {};
		\node [style=none] (2) at (1, 0.5) {$\frac{4 \pi}{3},0$};
	\end{pgfonlayer}
	\begin{pgfonlayer}{edgelayer}
		\draw (0) to (1.center);
	\end{pgfonlayer}
\end{tikzpicture}; \hspace{5pt} [\![ (xz^2)_2 ]\!]\equiv  \ket{\rightthreetimes}= \begin{tikzpicture}[scale=0.65]
	\begin{pgfonlayer}{nodelayer}
		\node [style=gn] (0) at (0, 0.5) {};
		\node [style=none] (1) at (0, -0.5) {};
		\node [style=none] (2) at (1, 0.5) {$0,\frac{4 \pi}{3}$};
	\end{pgfonlayer}
	\begin{pgfonlayer}{edgelayer}
		\draw (0) to (1.center);
	\end{pgfonlayer}
\end{tikzpicture}
\end{split}
\end{equation}

The post-selected measurements (indicator functions) are the adjoints of these and the reversible transformations in \textbf{3Spek} are mapped to the reversible of \textbf{3Stab} corresponding to the same spider diagram.
This means that our previous discussion of qutrit stabilizer quantum theory carries through to $\textbf{3Spek}$.
\end{proof}

In future, we can find the phase group of Spekkens-Schreiber theory for dits in any dimension D. This should allow us to depict these theories using (a version of) the qudit ZX calculus. We would then be able to study the relationship between Spekkens-Schreiber theory for dits and qudit stabilizer theory in the general case.
We will return to address this question and the issue of analyzing mutually unbiased qudit theories with arbitrary Abelian phase groups in another article.

\section{Further work}

We will conclude this paper with a brief outline of possible avenues of research which follow from the work presented here.
As we mentioned earlier, understanding how the qubit calculus fits into the general qudit calculus and proving the completeness of the generalized qudit ZX calculus for stabilizer quantum mechanics would certainly provide new insights into qudit stabilizer quantum mechanics. This might lead to modifications of the qudit ZX calculus before it reaches its final form.

For example, the relation between phase group structure and graph structure is still unclear. Can the qudit ZX calculus be expressed without angles by adding axioms relating to graph structure \cite{Per13}? It might be interesting to analyze alternative calculi with multiple edges between vertices. This approach could simplify proofs of completeness or provide a graphical depiction of non-locality.

Another possible mathematical framework for studying the qudit ZX calculus would be to use product and permutation categories (PROPs). This approach may allow an elegant synthetic axiomatization of numerous physical process theories and could provide new completeness theorems for corresponding graphical calculi.  

On a more practical note, the calculus for qudit stabilizer quantum theory can help describe quantum information protocols in a new light. One may generalize qubit protocols to qudits and try to understand new features of familiar quantum processes. For example, the formalism could be used to give a general description of error correction and fault tolerance for qudits. This could then be related to qubit error correction and links might be made between error correction in various dimensions. Furthermore, getting new insights into the abstract structure of qudit quantum mechanics could play a pivotal role in the development of new quantum algorithms. 

There are also a number of quantum foundations questions which could be addressed next. For instance, we know that the single qudit stabilizer theory is operationally  equivalent to Spekkens-Schreiber theory for dits for \textit{finite odd dimensions} and therefore admits a non-contextual, local hidden variable model in those cases. But what is the relationship between qudit stabilizer theory and Spekkens-Schreiber toy theory in general? We could also study van Enk's toy model\cite{Enk07} as a MUQT and find its phase group.

More generally, it would be useful to classify all the mutually unbiased qudit theories and determine which quantum-like features each one exhibits. For example, what is the relationship between a theory's phase group and whether it admits a local hidden variable interpretation? The study of the qudit ZX calculus with different Abelian phase groups should produce a large class of interesting toy models. In the future, we could also consider theories where distinct observable structures have different phase groups.

Moreover, the qubit ZX calculus, by providing a clear description of complementarity and phase, has helped clarify the relationship between complementarity and non-locality\cite{Co12}. The qudit ZX calculus provides the ideal framework to study other similar foundational questions related to complementarity. We could, for example, use the categorical framework to study how various notions of complementarity arise in different dimensions. Can one find a pictorial calculus (say a ZXY calculus for qudits) which captures complementarity of more than two observables? 

Finally, it would be interesting to understand the interpretation of the D-torus phase groups for qudit quantum mechanics observables from a physical point of view. What properties of the quantum phase group lead to specific quantum features? Perhaps studying the operational interpretation of phase \cite{Gard13} in physical theories could help us find the physical reason for each phase group taking the form it does. The study of phase and complementarity from an operational point of view may also provide insight into the relationship between categorical quantum mechanics and generalized probabilistic theories.

\section*{Acknowledgments}

I would like to thank both of my supervisors Bob Coecke and Terry Rudolph for insightful discussions. I am also very grateful to Mihai Vidrighin for his help with the illustrations and for his valuable comments. This research is funded by the EPSRC.

\begin{appendices}

\section{Derivation of the qudit ZX calculus}

A symmetric monoidal category (SMC) consists of a category C, a bifunctor  
$-\otimes$-: $C \times C \rightarrow C$, a unit object I and natural isomorphisms (with coherence conditions\cite{Mac98}): 
\begin{equation*}
\lambda_A: A \cong I \otimes A, \; \rho_A: A \cong A \otimes I, \; \alpha_{A,B,C}: A \otimes (B \otimes C) \cong (A \otimes B) \otimes C, \; \sigma_{A,B}: A \otimes B \cong B \otimes A
\end{equation*}

Monoidal categories are ideal for describing very general compositional theories of systems and processes \cite{Ab04}, since they contain two interacting modes $\otimes$ and $\circ$ of composition. These lead to a very simple diagrammatic calculus \cite{Sel09} where arrows are represented by boxes and the objects are vertical inputs/outputs. The $\otimes$ and $\circ$ operations are respectively represented as boxes juxtaposed next to each other and attached in vertical sequence. 

A dagger compact symmetric monoidal category ($\dagger$-CSMC) C is a SMC with an identity-on-objects contravariant dagger functor $\dagger:C \rightarrow C$ such that: 
\begin{equation*}
 (f \circ g)^{\dagger}=g^{\dagger} \circ f^{\dagger}, \; (f \otimes g)^{\dagger}=f^{\dagger} \otimes g^{\dagger}, \; id_A^{\dagger}=id_A, \; (f^{\dagger})^{\dagger}=f     
 \end{equation*}

which is also compact, meaning that each object $A \in \vert C \vert $ has a dual object $\bar{A} \in \vert C \vert $ (note that we will always use $A=\bar{A}$) and arrows: $\eta_A: I \rightarrow \bar{A} \otimes A$ and $\epsilon_A: A \otimes \bar{A} \rightarrow I$ such that: 
\begin{equation*}
(\epsilon_A \otimes id_A) \circ (id_A \otimes \eta_A)= id_A \text{ and } (id_{\bar{A}} \otimes \epsilon_A ) \circ (\eta_A \otimes \bar{A})= id_{\bar{A}} \\
\end{equation*} 

We define a state of a system A as a morphism: $\psi: I \rightarrow A$, an effect as: $\pi: A \rightarrow I$ and scalars as $s: I \rightarrow I$. The inner product between states is then the scalar: $\psi^{\dagger} \circ \phi: I \rightarrow I$. The dimension of an object A is defined as: 
$dim(A):= \eta_A^\dagger \circ \eta_A$.

If we add to the previous graphical calculus a vertical involutive asymmetry in the boxes representing arrows and the rule that taking the adjoint reflects the boxes vertically, then we get the following key theorem which allows us to use graphical reasoning:

\underline{Theorem A1:} An equational statement between formal expressions in the language of $\dagger$-CSMC holds if and only if it holds up to isotopy in the graphical calculus.  \cite{Kel80,Sel07} 

Important examples of $\dagger$-CSMCs are \textbf{FHilb}, the category of finite dimensional Hilbert spaces and bounded linear maps with the usual tensor product, and \textbf{FRel}, the category of finite sets and relations with the Cartesian product of sets as the tensor product.

We can describe bases and observables in the general context of $\dagger$-CSMC by noting that the contrapositive of the no cloning \cite{Wo82} and no deleting theorems \cite{Pa00} states that orthonormal basis states are the only ones which can be copied and erased. 

An observable structure is a $\dagger$-special commutative Frobenius algebra on a $\dagger$-CSMC C \cite{Pa08}. This is a triple 
\begin{center}
$\{$ A $\in$ $\vert C \vert$,  
$\delta:A\rightarrow A \otimes A$ \begin{tikzpicture}[scale=0.8]
	\begin{pgfonlayer}{nodelayer}
		\node [style=none] (0) at (0, 1) {};
		\node [style=none] (1) at (0.4999993, -1) {};
		\node [style=none] (2) at (-0.4999993, -1) {};
		\node [style=gn] (3) at (0, -0) {};
	\end{pgfonlayer}
	\begin{pgfonlayer}{edgelayer}
		\draw (0.center) to (3);
		\draw (3) to (1.center);
		\draw (3) to (2.center);
	\end{pgfonlayer}
\end{tikzpicture} (copying map),  
$\epsilon:A\rightarrow I$ \begin{tikzpicture}[scale=0.8]
	\begin{pgfonlayer}{nodelayer}
		\node [style=none] (0) at (0, 0.5) {};
		\node [style=gn] (1) at (0, -0.5) {};
	\end{pgfonlayer}
	\begin{pgfonlayer}{edgelayer}
		\draw (0.center) to (1);
	\end{pgfonlayer}
\end{tikzpicture} (erasing map) $\}$ \\
\end{center} 
satisfying: 
\begin{equation*}
(\delta \otimes id_{A}) \circ \delta= ( id_{A} \otimes\delta ) \circ \delta; \; \lambda_A^{-1} \circ (\epsilon \otimes id_{A}) \circ \ \delta=\rho_A^{-1} \circ ( id_{A} \otimes \epsilon ) \circ \delta= id_A; \; \sigma_{A,A} \circ \delta=\delta; \;
\end{equation*} 
\begin{equation*}
(\delta^{\dagger} \otimes id_A) \circ (id_A \otimes \delta)= \delta \circ \delta^{\dagger}; \; \delta^{\dagger} \circ \delta = id_A 
\end{equation*}

In \textbf{FHilb}, orthonormal bases are in a one to one correspondence with $\dagger$-special commutative Frobenius algebras \cite{Pav07}. This definition for observable structures has been shown \cite{Paq06} to be equivalent to the spider laws depicted below.

\begin{center}
\begin{tikzpicture}[scale=0.5]
	\begin{pgfonlayer}{nodelayer}
		\node [style=rn] (0) at (-11, -1) {};
		\node [style=none] (1) at (-14, 2) {};
		\node [style=none] (2) at (-12, 2) {};
		\node [style=none] (3) at (-12, -2) {};
		\node [style=none] (4) at (-10, -2) {};
		\node [style=none] (5) at (-11, -2) {...};
		\node [style=none] (6) at (-12, -0) {...};
		\node [style=none] (7) at (-13, 2) {...};
		\node [style=none] (8) at (-7.999999, -0) {=};
		\node [style=rn] (9) at (-5, -0) {};
		\node [style=none] (10) at (-6, 2) {};
		\node [style=none] (11) at (-4, 2) {};
		\node [style=none] (12) at (-6, -2) {};
		\node [style=none] (13) at (-4, -2) {};
		\node [style=none] (14) at (-14.5, -2) {};
		\node [style=none] (15) at (-13, -2) {};
		\node [style=none] (16) at (-11, 2) {};
		\node [style=none] (17) at (-9.500001, 2) {};
		\node [style=none] (18) at (-10.25, 2) {...};
		\node [style=none] (19) at (-13.75, -2) {...};
		\node [style=none] (20) at (-5, -2) {...};
		\node [style=none] (21) at (-5, 2) {...};
		\node [style=none] (22) at (0, -0) {;};
		\node [style=none] (23) at (2.999999, -2) {};
		\node [style=rn] (24) at (2.999999, -0) {};
		\node [style=none] (25) at (5.999999, -2) {};
		\node [style=none] (26) at (2.999999, 2) {};
		\node [style=none] (27) at (5.999999, 2) {};
		\node [style=none] (28) at (4.5, -0) {=};
		\node [style=rn] (30) at (-13, 1) {};
		\node [style=gn] (31) at (-13, 1) {};
		\node [style=gn] (32) at (-11, -1) {};
		\node [style=gn] (33) at (-5, -0) {};
		\node [style=gn] (34) at (2.999999, -0) {};
	\end{pgfonlayer}
	\begin{pgfonlayer}{edgelayer}
		\draw [style=none] (0) to (3.center);
		\draw [style=none] (0) to (4.center);
		\draw [style=none] (10.center) to (9);
		\draw [style=none] (9) to (11.center);
		\draw [style=none] (9) to (13.center);
		\draw [style=none] (9) to (12.center);
		\draw [style=none, bend right=15, looseness=1.00] (0) to (16.center);
		\draw [style=none] (0) to (17.center);
		\draw [style=none] (26.center) to (24);
		\draw [style=none] (24) to (23.center);
		\draw [style=none] (27.center) to (25.center);
		\draw [style=none, bend left=45, looseness=1.25] (30) to (0);
		\draw [style=none, bend right=45, looseness=1.50] (30) to (0);
		\draw [style=none, bend right=15, looseness=1.00] (30) to (15.center);
		\draw (2.center) to (30);
		\draw [style=none] (30) to (14.center);
		\draw (1.center) to (30);
	\end{pgfonlayer}
\end{tikzpicture}
\end{center}

We define a classical point for an observable structure (A, $\delta$, $\epsilon$) as a self conjugate morphism \\ 
k: $I \rightarrow A$ 
\begin{tikzpicture}[scale=0.8]
	\begin{pgfonlayer}{nodelayer}
		\node [style=rn] (0) at (0, 0.5) {k};
		\node [style=none] (1) at (0, -0.5) {};
	\end{pgfonlayer}
	\begin{pgfonlayer}{edgelayer}
		\draw (0) to (1.center);
	\end{pgfonlayer}
\end{tikzpicture} obeying: \\
\begin{center}
\begin{tikzpicture}[scale=0.6]
	\begin{pgfonlayer}{nodelayer}
		\node [style=none] (0) at (-1, -0) {};
		\node [style=gn] (1) at (-1, -0) {};
		\node [style=none] (2) at (0, -0) {=};
		\node [style=none] (3) at (-1.5, -1) {};
		\node [style=none] (4) at (-0.5, -1) {};
		\node [style=none] (5) at (1, -0.5) {};
		\node [style=none] (6) at (2, -0.5) {};
		\node [style=rn] (7) at (2, 0.5) {k};
		\node [style=rn] (8) at (1, 0.5) {k};
		\node [style=rn] (9) at (-1, 1) {k};
	\end{pgfonlayer}
	\begin{pgfonlayer}{edgelayer}
		\draw (0.center) to (3.center);
		\draw (0.center) to (4.center);
		\draw (8) to (5.center);
		\draw (7) to (6.center);
		\draw (9) to (0.center);
	\end{pgfonlayer}
\end{tikzpicture}  \hspace{50pt}  and  \hspace{50pt}  
\begin{tikzpicture}
	\begin{pgfonlayer}{nodelayer}
		\node [style=none] (0) at (-1, -0.5) {};
		\node [style=gn] (1) at (-1, -0.5) {};
		\node [style=none] (2) at (0, -0) {=};
		\node [style=rn] (3) at (-1, 0.5) {k};
	\end{pgfonlayer}
	\begin{pgfonlayer}{edgelayer}
		\draw (3) to (0.center);
	\end{pgfonlayer}
\end{tikzpicture}
\end{center}

This means that classical points are those which get copied by the copying map and deleted by the deleting map. In \textbf{FHilb}, for example, they are the basis states corresponding to the observable structure. \\

We will now introduce a notion of phase relative to a given basis \cite{Du08} which allows us to study unbiasedness and the interplay between several bases.  

Let (A, $\delta$, $\epsilon$) be an observable structure. For any two points $\psi$, $\phi$: $I \rightarrow A$, we define a multiplication operation:
\begin{equation}
\psi \odot \phi= \delta^{\dagger} \circ (\psi \otimes \phi) \circ \lambda_I 
\end{equation} 

Note that this multiplication on points is commutative, associative and $\epsilon^{\dagger} \odot \psi =\psi$ for any point $\psi$.

A point $\alpha: I \rightarrow A $ is called \textit{unbiased} relative to an observable structure (A, $\delta$, $\epsilon$) if there exists a scalar s: $I \rightarrow I$ such that: $s. \alpha \odot \alpha^{\star}= \epsilon^{\dagger}$. This is a generalization of the usual definition of an unbiased vector with respect to a basis.

For each state and observable structure (A, $\delta$, $\epsilon$), we introduce a \textit{phase map} $\Lambda$ which takes each point $\psi$: $I \rightarrow A$, to the morphism: 
\begin{equation}
\Lambda (\psi)= \delta^{\dagger} \circ (\psi \otimes 1_A): A \rightarrow A
\end{equation} 

The phase map satisfies several properties: \\
(i)$\Lambda(\psi \odot \phi)= \Lambda(\psi) \circ \Lambda(\phi)$
(ii) $\Lambda(\epsilon^{\dagger})=1_A$
(iii) $\Lambda(\psi)=\Lambda(\psi)^{\dagger}$
(iv) Phase maps commute freely with the observable structure since: 
$\Lambda(\psi) \circ \delta^{\dagger} = \delta^{\dagger} \circ (1_A \otimes \Lambda(\psi)) = \delta^{\dagger} \circ (\Lambda(\psi) \otimes 1_A)$.

We can extend the spider laws to account for phases relative to an observable structure.

\underline{Theorem A2:} Any morphism $A^{\otimes n} \rightarrow A^{\otimes m}$ generated from an observable structure (A, $\delta$, $\epsilon$), together with one occurrence of each unbiased point $\alpha_i:$ $I \rightarrow A$ can be written in the form \cite{Du08}:
\begin{center}
\begin{tikzpicture}[scale=0.4]
	\begin{pgfonlayer}{nodelayer}
		\node [style=gn] (0) at (4, -0) {};
		\node [style=gn] (1) at (4, 1) {};
		\node [style=gn] (2) at (3, 2) {};
		\node [style=gn] (3) at (2, 3) {};
		\node [style=gn] (4) at (4, -1) {};
		\node [style=gn] (5) at (5, -2) {};
		\node [style=gn] (6) at (6, -3) {};
		\node [style=none] (7) at (5, -4) {};
		\node [style=none] (8) at (3, -4) {};
		\node [style=none] (9) at (7, -4) {};
		\node [style=none] (10) at (4, -4) {...};
		\node [style=none] (11) at (1, 4) {};
		\node [style=none] (12) at (1, -4) {};
		\node [style=none] (13) at (3, 4) {};
		\node [style=none] (14) at (4, 4) {...};
		\node [style=none] (15) at (5, 4) {};
		\node [style=none] (16) at (7, 4) {};
		\node [style=none] (17) at (0, -0) {:=};
		\node [style=gn] (18) at (-4, -0) {};
		\node [style=none] (19) at (-1, 4) {};
		\node [style=none] (20) at (-7, 4) {};
		\node [style=none] (21) at (-5.5, 4) {};
		\node [style=none] (22) at (-4, 4) {...};
		\node [style=none] (23) at (-2.5, 4) {};
		\node [style=none] (24) at (-1, -4) {};
		\node [style=none] (25) at (-2.5, -4) {};
		\node [style=none] (26) at (-4, -4) {...};
		\node [style=none] (27) at (-5.5, -4) {};
		\node [style=none] (28) at (-7, -4) {};
		\node [style=none] (29) at (-2.5, -0) {$\bigodot_i \alpha_i$};
		\node [style=none] (30) at (5.5, -0) {$\bigodot_i \alpha_i$};
	\end{pgfonlayer}
	\begin{pgfonlayer}{edgelayer}
		\draw (20.center) to (18);
		\draw (21.center) to (18);
		\draw (23.center) to (18);
		\draw (19.center) to (18);
		\draw (18) to (24.center);
		\draw (18) to (25.center);
		\draw (18) to (27.center);
		\draw (18) to (28.center);
		\draw (11.center) to (3);
		\draw (13.center) to (3);
		\draw (2) to (3);
		\draw (2) to (15.center);
		\draw (2) to (1);
		\draw (1) to (16.center);
		\draw (1) to (0);
		\draw (0) to (4);
		\draw (4) to (5);
		\draw (5) to (8.center);
		\draw (4) to (12.center);
		\draw (6) to (5);
		\draw (6) to (7.center);
		\draw (6) to (9.center);
	\end{pgfonlayer}
\end{tikzpicture} where: \hspace{10pt} \begin{tikzpicture}[scale=0.7]
	\begin{pgfonlayer}{nodelayer}
		\node [style=gn] (0) at (0, -0) {};
		\node [style=none] (1) at (0, 1) {};
		\node [style=none] (2) at (0, -1) {};
		\node [style=none] (3) at (1, -0) {$\bigodot_i \alpha_i$};
	\end{pgfonlayer}
	\begin{pgfonlayer}{edgelayer}
		\draw (1.center) to (0);
		\draw (0) to (2.center);
	\end{pgfonlayer}
\end{tikzpicture} \hspace{6pt} is the phase map $\Lambda (\bigodot_i \alpha_i)$.
\end{center}

These spider maps compose according to the generalized spider law:

\begin{center}
\begin{tikzpicture}[scale=0.6]
	\begin{pgfonlayer}{nodelayer}
		\node [style=none] (0) at (-6, 2) {};
		\node [style=none] (1) at (-4, 2) {};
		\node [style=none] (2) at (-4, -2) {};
		\node [style=none] (3) at (-2, -2) {};
		\node [style=none] (4) at (-3, -2) {...};
		\node [style=none] (5) at (-4, -0) {...};
		\node [style=none] (6) at (-5, 2) {...};
		\node [style=none] (7) at (0, -0) {=};
		\node [style=rn] (8) at (3, -0) {};
		\node [style=none] (9) at (2, 2) {};
		\node [style=none] (10) at (4, 2) {};
		\node [style=none] (11) at (2, -2) {};
		\node [style=none] (12) at (4, -2) {};
		\node [style=none] (13) at (-6.5, -2) {};
		\node [style=none] (14) at (-5, -2) {};
		\node [style=none] (15) at (-3, 2) {};
		\node [style=none] (16) at (-1.5, 2) {};
		\node [style=none] (17) at (-2.25, 2) {...};
		\node [style=none] (18) at (-5.75, -2) {...};
		\node [style=none] (19) at (3, -2) {...};
		\node [style=none] (20) at (3, 2) {...};
		\node [style=none] (21) at (4.5, -0) {$\alpha \odot \beta$};
		\node [style=gn] (22) at (3, -0) {};
		\node [style=rn] (23) at (-3, -0.9999995) {};
		\node [style=rn] (24) at (-5, 1) {};
		\node [style=gn] (25) at (-5, 0.9999998) {};
		\node [style=gn] (26) at (-3, -0.9999998) {};
		\node [style=gn] (27) at (-5, 0.9999998) {};
		\node [style=gn] (28) at (-3, -0.9999998) {};
		\node [style=none] (29) at (-2.25, -0.9999998) {$\beta$};
		\node [style=none] (30) at (-5.75, 0.9999998) {$\alpha$};
	\end{pgfonlayer}
	\begin{pgfonlayer}{edgelayer}
		\draw [style=none] (9.center) to (8);
		\draw [style=none] (8) to (10.center);
		\draw [style=none] (8) to (12.center);
		\draw [style=none] (8) to (11.center);
		\draw [style=none] (23) to (3.center);
		\draw [style=none] (23) to (2.center);
		\draw [style=none] (23) to (16.center);
		\draw [style=none, bend right=15, looseness=1.00] (23) to (15.center);
		\draw [style=none, bend left=45, looseness=1.25] (24) to (23);
		\draw [style=none] (24) to (13.center);
		\draw [style=none, bend right=45, looseness=1.50] (24) to (23);
		\draw (1.center) to (24);
		\draw (0.center) to (24);
		\draw [style=none, bend right=15, looseness=1.00] (24) to (14.center);
	\end{pgfonlayer}
\end{tikzpicture}.
\end{center}

This theorem follows from the spider laws together with the fact that the phase maps commute freely with the observable structure (property (iv) given above).

Let $\alpha$: $I \rightarrow A$ be a point satisfying $\alpha^{\dagger} \circ \alpha=dim(A)$. Then $\alpha$ is unbiased iff $\Lambda(\alpha)$ is unitary. \\
Note that the choice of $\alpha^{\dagger} \circ \alpha=dim(A)$ is taken for unbiased points from this point onwards. \\

All the points which are unbiased with respect to the basis corresponding to the observable structure (A,$\delta$, $\epsilon$) form an Abelian group $\mathcal{U}$ with respect to the multiplication $\odot$. This is clear since $\odot$ is closed for unbiased points, commutative, associative, admits the unique identity point $\epsilon^{\dagger}$ and each point has a unique inverse, its conjugate. \\

The phase maps, restricted to act on unbiased points relative to the observable structure, form an abelian group with map composition as the group operation, which is isomorphic to $\mathcal{U}$ . We call this the \textbf{\textit{phase group}} $\Pi$. \\

Note that, for each unbiased point $\alpha$ we can define a new observable structure (A, $\delta_{\alpha}$, $\epsilon_{\alpha}$) where: \\
\begin{center}
$\delta_{\alpha}$=\begin{tikzpicture}[scale=0.7]
	\begin{pgfonlayer}{nodelayer}
		\node [style=gn] (0) at (0, -0) {$\alpha$};
		\node [style=none] (1) at (0.9999998, -0.9999998) {};
		\node [style=none] (2) at (0, 0.9999998) {};
		\node [style=none] (3) at (-0.9999998, -0.9999998) {};
	\end{pgfonlayer}
	\begin{pgfonlayer}{edgelayer}
		\draw (2.center) to (0);
		\draw (0) to (1.center);
		\draw (0) to (3.center);
	\end{pgfonlayer}
\end{tikzpicture} := $(\Lambda(\alpha) \otimes \Lambda(\alpha)) \circ \delta \circ \Lambda(\alpha)^{\dagger}$  and  $\epsilon_{\alpha}$= \begin{tikzpicture}[scale=0.7]
	\begin{pgfonlayer}{nodelayer}
		\node [style=none] (0) at (0, 0.5) {};
		\node [style=gn] (1) at (0, -0.5) {$\alpha$};
	\end{pgfonlayer}
	\begin{pgfonlayer}{edgelayer}
		\draw (0.center) to (1);
	\end{pgfonlayer}
\end{tikzpicture} := $\epsilon \circ \Lambda(\alpha)^{\dagger}$.
\end{center}

We use the phase group for observable structures as a tool to study physical theories from an abstract algebraic perspective.

We now proceed to study how two complementary observable structures interact\cite{Du08}. In general, we cannot assume that the dagger compact structures of two distinct observable structures coincide \cite{Coe08}. 
 
 Therefore, we define the \textit{dualizer} of observable structures $(A,\delta_Z,\epsilon_Z)$ and $(A,\delta_X,\epsilon_X)$ as: \\
\begin{center}
\begin{tikzpicture}[scale=0.5]
	\begin{pgfonlayer}{nodelayer}
		\node [style=rn] (0) at (3, 1) {};
		\node [style=none] (1) at (4.75, -1.25) {};
		\node [style=none] (2) at (2, -0) {};
		\node [style=none] (3) at (-1.5, -0) {:=};
		\node [style=gn] (4) at (1, -1) {};
		\node [style=none] (5) at (-0.75, 1.25) {};
		\node [style=none] (6) at (-2.5, 1) {};
		\node [style=none] (7) at (-2.5, -1) {};
		\node [style=Had] (8) at (-2.5, -0) {S};
	\end{pgfonlayer}
	\begin{pgfonlayer}{edgelayer}
		\draw [bend left, looseness=1.00] (0) to (1.center);
		\draw [bend right, looseness=1.00] (0) to (2.center);
		\draw [bend left=15, looseness=1.25] (2.center) to (4);
		\draw [bend right, looseness=0.75] (5.center) to (4);
		\draw (6.center) to (8);
		\draw (8) to (7.center);
	\end{pgfonlayer}
\end{tikzpicture}
\end{center}
 
By dagger compactness, we can see that the dualizer is unitary. This shows that the dimension of a dagger symmetric monoidal category does not depend on the choice of observable structure\cite{Du08} since: \\
\begin{center}
dim(A):= \begin{tikzpicture}[scale=0.5]
	\begin{pgfonlayer}{nodelayer}
		\node [style=gn] (0) at (-2, 1) {};
		\node [style=rn] (1) at (6, 1) {};
		\node [style=none] (2) at (5, -0) {};
		\node [style=none] (3) at (7, -0) {};
		\node [style=none] (4) at (-1, -0) {};
		\node [style=none] (5) at (-3, -0) {};
		\node [style=none] (6) at (0, -0) {=};
		\node [style=none] (7) at (-4, -0) {=};
		\node [style=none] (8) at (-7, -0) {};
		\node [style=none] (9) at (-5, -0) {};
		\node [style=none] (10) at (-6, 1) {};
		\node [style=none] (11) at (-6, -1) {};
		\node [style=gn] (12) at (-2, -1) {};
		\node [style=rn] (13) at (2, -2) {};
		\node [style=rn] (14) at (2, 2) {};
		\node [style=rn] (15) at (6, -1) {};
		\node [style=none] (16) at (1, 1) {};
		\node [style=none] (17) at (3, 1) {};
		\node [style=none] (18) at (4, -0) {=};
		\node [style=Had] (19) at (1, 0.5) {$S^{\dagger}$};
		\node [style=Had] (20) at (1, -0.5) {S};
		\node [style=none] (21) at (1, -1) {};
		\node [style=none] (22) at (3, -1) {};
	\end{pgfonlayer}
	\begin{pgfonlayer}{edgelayer}
		\draw [bend left=45, looseness=1.25] (5.center) to (0);
		\draw [in=90, out=0, looseness=1.25] (0) to (4.center);
		\draw [bend left=45, looseness=1.25] (2.center) to (1);
		\draw [in=90, out=0, looseness=1.25] (1) to (3.center);
		\draw [bend left=45, looseness=1.25] (16.center) to (14);
		\draw [in=90, out=0, looseness=1.25] (14) to (17.center);
		\draw [in=180, out=90, looseness=1.25] (8.center) to (10.center);
		\draw [bend left=45, looseness=1.25] (10.center) to (9.center);
		\draw [bend left=45, looseness=1.00] (9.center) to (11.center);
		\draw [in=-90, out=180, looseness=1.25] (11.center) to (8.center);
		\draw [bend right=45, looseness=1.00] (5.center) to (12);
		\draw [bend right=45, looseness=1.25] (12) to (4.center);
		\draw (16.center) to (19);
		\draw [bend left=45, looseness=1.25] (22.center) to (13);
		\draw [bend right=45, looseness=1.00] (21.center) to (13);
		\draw (21.center) to (20);
		\draw (20) to (19);
		\draw [bend right=45, looseness=1.25] (2.center) to (15);
		\draw [bend right=45, looseness=1.25] (15) to (3.center);
		\draw (17.center) to (22.center);
	\end{pgfonlayer}
\end{tikzpicture}
\end{center}

In a Hilbert space of D dimensions, two orthonormal bases $\{u_1, u_2, ..., u_D \}$  and $\{ v_1, v_2, ..., v_D \}$ are called \textit{unbiased} if:
\begin{equation}
D \left| \langle v_i , u_j \rangle \right|^2 = 1, \forall i,j  \{1, 2, ..., D \}
\end{equation}

If a quantum system is prepared in a state corresponding to a vector in one of these bases, then all the outcomes of a measurement with respect to the other mutually unbiased basis will occur with equal probabilities. No information can be retrieved by performing such a measurement. In this sense, two mutually unbiased bases corresponding to eigenstates of two non-degenerate quantum observables describe mutually exclusive physical measurement procedures. 

This provides a mathematical expression for Bohr's principle of complementarity that: ``evidence obtained under different experimental conditions cannot be comprehended within a single picture, but must be regarded as complementary in the sense that only the totality of the phenomena exhausts the possible information about the objects".
 
Two observable structures $(A,\delta_Z,\epsilon_Z)$ and $(A,\delta_X,\epsilon_X)$ are called \textit{\textbf{complementary}} if: \\
(i) whenever a point z: $I \rightarrow A$ \begin{tikzpicture}[scale=0.8]
	\begin{pgfonlayer}{nodelayer}
		\node [style=rn] (0) at (0, 0.5) {z};
		\node [style=none] (1) at (0, -0.5) {};
	\end{pgfonlayer}
	\begin{pgfonlayer}{edgelayer}
		\draw (0) to (1.center);
	\end{pgfonlayer}
\end{tikzpicture} is classical for $(\delta_Z,\epsilon_Z)$ it is unbiased for $(\delta_X,\epsilon_X))$. \\
(ii) whenever a point x: $I \rightarrow A$ \begin{tikzpicture}[scale=0.8]
	\begin{pgfonlayer}{nodelayer}
		\node [style=gn] (0) at (0, 0.5) {x};
		\node [style=none] (1) at (0, -0.5) {};
	\end{pgfonlayer}
	\begin{pgfonlayer}{edgelayer}
		\draw (0) to (1.center);
	\end{pgfonlayer}
\end{tikzpicture} is classical for $(\delta_X,\epsilon_X)$ it is unbiased for $(\delta_Z,\epsilon_Z))$. 

This definition of complementarity could easily be generalized to more than two observable structures by requiring that whenever a point is classical for one observable structure, it must be unbiased for all the other observable structures.
One can show the following theorem\cite{Du08}, assuming that at least one of the observable structures describes a basis:

\underline{Theorem A3:} Two observable structures are complementary iff they obey: \\ 
\begin{center}
\begin{tikzpicture}[scale=0.5]
	\begin{pgfonlayer}{nodelayer}
		\node [style=gn] (0) at (-1.5, 1) {};
		\node [style=rn] (1) at (-1.5, -1) {};
		\node [style=none] (2) at (-1.5, 2) {};
		\node [style=none] (3) at (-1.5, -2) {};
		\node [style=none] (4) at (-3.5, -0) {};
		\node [style=none] (5) at (-4.5, -0) {};
		\node [style=none] (6) at (-4, 0.5) {};
		\node [style=none] (7) at (-4, -0.5) {};
		\node [style=none] (8) at (1.5, -2) {};
		\node [style=none] (9) at (1.5, 2) {};
		\node [style=gn] (10) at (1.5, 1) {};
		\node [style=rn] (11) at (1.5, -1) {};
		\node [style=Had] (12) at (-2, -0) {S};
		\node [style=none] (13) at (0, -0) {=};
	\end{pgfonlayer}
	\begin{pgfonlayer}{edgelayer}
		\draw (9.center) to (10);
		\draw (11) to (8.center);
		\draw (3.center) to (1);
		\draw [bend left, looseness=1.00] (1) to (12);
		\draw [bend left, looseness=1.00] (12) to (0);
		\draw [bend left=45, looseness=1.25] (0) to (1);
		\draw [bend right=45, looseness=1.25] (4.center) to (6.center);
		\draw (0) to (2.center);
		\draw [bend left=45, looseness=1.00] (4.center) to (7.center);
		\draw [bend right=45, looseness=1.00] (5.center) to (7.center);
		\draw [bend left=45, looseness=1.00] (5.center) to (6.center);
	\end{pgfonlayer}
\end{tikzpicture}
\end{center}

Two observable structures $(A,\delta_Z,\epsilon_Z)$ and $(A,\delta_X,\epsilon_X)$ are called \textit{coherent} if the erasing point for each observable structure is a classical point for the other observable structure. This can be pictured as: \\
\begin{center}
\begin{tikzpicture}[scale=0.7]
	\begin{pgfonlayer}{nodelayer}
		\node [style=rn] (0) at (-6, 0.5) {};
		\node [style=gn] (1) at (-6, -0.5) {};
		\node [style=rn] (2) at (-4.5, 1) {};
		\node [style=gn] (3) at (-4.5, -0) {};
		\node [style=none] (4) at (-5, -1) {};
		\node [style=none] (5) at (-4, -1) {};
		\node [style=rn] (6) at (-2, 0.5) {};
		\node [style=rn] (7) at (-1, 0.5) {};
		\node [style=none] (8) at (-2, -0.5) {};
		\node [style=none] (9) at (-3, -0) {=};
		\node [style=none] (10) at (-1, -0.5) {};
		\node [style=none] (11) at (0, -0) {and};
		\node [style=gn] (12) at (1, 0.5) {};
		\node [style=rn] (13) at (1, -0.5) {};
		\node [style=rn] (14) at (2.5, -0) {};
		\node [style=gn] (15) at (2.5, 1) {};
		\node [style=gn] (16) at (5, 0.5) {};
		\node [style=gn] (17) at (6, 0.5) {};
		\node [style=none] (18) at (3, -1) {};
		\node [style=none] (19) at (4, -0) {=};
		\node [style=none] (20) at (5, -0.5) {};
		\node [style=none] (21) at (6, -0.5) {};
		\node [style=none] (22) at (2, -1) {};
	\end{pgfonlayer}
	\begin{pgfonlayer}{edgelayer}
		\draw (0) to (1);
		\draw (2) to (3);
		\draw (3) to (4.center);
		\draw (3) to (5.center);
		\draw (6) to (8.center);
		\draw (7) to (10.center);
		\draw (12) to (13);
		\draw (15) to (14);
		\draw (14) to (22.center);
		\draw (14) to (18.center);
		\draw (16) to (20.center);
		\draw (17) to (21.center);
	\end{pgfonlayer}
\end{tikzpicture}
\end{center}

In line with qudit quantum theory, states and erasing points are defined such that: \\
\begin{center}
\begin{tikzpicture}[scale=0.5]
	\begin{pgfonlayer}{nodelayer}
		\node [style=none] (0) at (0, -0) {=};
		\node [style=none] (1) at (-3, -0) {};
		\node [style=none] (2) at (-1, -0) {};
		\node [style=none] (3) at (-2, 1) {};
		\node [style=none] (4) at (-2, -1) {};
		\node [style=gn] (5) at (1, 1) {};
		\node [style=gn] (6) at (2.5, -1) {};
		\node [style=rn] (7) at (2.5, 1) {};
		\node [style=rn] (8) at (1, -1) {};
	\end{pgfonlayer}
	\begin{pgfonlayer}{edgelayer}
		\draw [in=180, out=90, looseness=1.25] (1.center) to (3.center);
		\draw [bend left=45, looseness=1.25] (3.center) to (2.center);
		\draw [bend left=45, looseness=1.00] (2.center) to (4.center);
		\draw [in=-90, out=180, looseness=1.25] (4.center) to (1.center);
		\draw (5) to (8);
		\draw (7) to (6);
	\end{pgfonlayer}
\end{tikzpicture}
\end{center}

The classical points $K_Z$ of an observable structure $\{A, \delta_Z, \epsilon_Z \}$ are called \textit{closed} for an observable structure $\{A, \delta_X, \epsilon_X \}$ if for all $k,k'\in K_Z$ we have $k \odot_X k'= \delta_X \circ (k \otimes k') \in K_Z$.

One can easily show\cite{Du08} that for every Hilbert space we can find a pair of coherent, closed, complementary observable structures. In fact, the observable structures corresponding to the Z and X qudit operators are closed, coherent and complementary.

Two observables structures $\{A, \delta_Z, \epsilon_Z \}$ and $\{A, \delta_X, \epsilon_X \}$ are said to be \textbf{\textit{strongly complementary}} if: 
\begin{center}
\begin{tikzpicture}[scale=0.7]
	\begin{pgfonlayer}{nodelayer}
		\node [style=none] (0) at (-2.5, 1.5) {};
		\node [style=none] (1) at (-1.5, 1.5) {};
		\node [style=gn] (2) at (-2.5, 0.4999998) {};
		\node [style=gn] (3) at (-1.5, 0.4999998) {};
		\node [style=rn] (4) at (-2.5, -0.5000003) {};
		\node [style=rn] (5) at (-1.5, -0.5000003) {};
		\node [style=none] (6) at (-2.5, -1.5) {};
		\node [style=none] (7) at (-1.5, -1.5) {};
		\node [style=none] (8) at (0, -0) {=};
		\node [style=none] (9) at (1.5, 1.5) {};
		\node [style=none] (10) at (2.5, 1.5) {};
		\node [style=none] (11) at (1.5, -1.5) {};
		\node [style=none] (12) at (2.5, -1.5) {};
		\node [style=gn] (13) at (2, -0.5000003) {};
		\node [style=rn] (14) at (2, 0.4999998) {};
		\node [style=rn] (15) at (-3.75, -0.5000001) {};
		\node [style=gn] (16) at (-3.75, 0.5000001) {};
	\end{pgfonlayer}
	\begin{pgfonlayer}{edgelayer}
		\draw (0.center) to (2);
		\draw (1.center) to (3);
		\draw (3) to (4);
		\draw (2) to (5);
		\draw (5) to (7.center);
		\draw (4) to (6.center);
		\draw (2) to (4);
		\draw (5) to (3);
		\draw (9.center) to (14);
		\draw (10.center) to (14);
		\draw (14) to (13);
		\draw (13) to (11.center);
		\draw (13) to (12.center);
		\draw (15) to (16);
	\end{pgfonlayer}
\end{tikzpicture} 
\end{center}

This condition is called strong complementarity since: \\
\underline{Theorem A4\cite{Du08}:} A pair of coherent, strongly complementary observable structures are complementary.  

One can show that:

\underline{Theorem A5\cite{Du08}:} If $\{A, \delta_Z, \epsilon_Z \}$ and $\{A, \delta_X, \epsilon_X \}$ are coherent strongly complementary observable structures and the set $K_X$ of classical points for $\{A, \delta_X, \epsilon_X \}$ is finite, then $K_X$ is a subgroup of the group $(U_Z, \odot_Z)$ of unbiased points for $\{A, \delta_Z, \epsilon_Z \}$.

The ZX calculus for qubits is restricted to two dimensions. However, since this algebraic characterization of bases applies to arbitrary dimensions, we can generalize the pictorial calculus to higher dimensional quantum systems. 

As with the qubit ZX calculus, the use of graphical notation is justified since \textbf{FHilb} is a $\dagger$-CSMC. We let all the edges be implicitly labeled by $\mathbb{C}^D$ and focus on a pair of observable structures corresponding to the Z and X observables from qudit quantum mechanics.

The green observable structure, corresponding to the qudit observable $Z=\sum_{j=0}^{D-1} \eta^j \ket{j}\bra{j}$ is defined via the copying and deleting maps:
\begin{center}
$\delta_Z =\left( \begin{array}{cccc}
e_1 |& e_2| & ... &| e_D   \end{array} \right)$  and $\epsilon_Z= \frac{1}{D}(1,1,..., 1)$. \\
\end{center} 

Where $\delta_Z$ is a $D^2 \times D$ matrix with D columns $e_i$ which have one 1 in row $D\times (i-1)+i$ and zeros in all the other rows. 

Unbiased points for the green observable structure are in the form: $\ket{\{\alpha_1, \alpha_2, ..., \alpha_{D-1} \}_Z}=\ket{0}+\sum_{j=1}^{D-1}e^{i \alpha_j} \ket{j}$ and therefore, the phase group consists of matrices of the form:
\begin{equation}
\Lambda_Z(\alpha_1, \alpha_2, ..., \alpha_{D-1})=\left( \begin{array}{ccccc}
1 & 0 & 0 & ... & 0  \\
0 & e^{i \alpha_1} & 0 & ... & 0 \\
0 & 0 & e^{i \alpha_1} & ... & ... \\
... & ... & ... & ... & 0 \\
0 & 0 & ... & 0 & e^{i \alpha_{D-1}} \end{array} \right)
\end{equation} 

Therefore the phase group for the green observable $\Pi_Z$ and the group of unbiased points $(U_Z, \odot_Z)$ are both the \textit{\textbf{D-torus group}}, corresponding to the direct product of D circle groups $S^1 \times ...\times S^1$. 

The green part of the ZX calculus for qudits follows from the generalized green spider law.

The red observable structure, corresponding to the qudit observable $X=\sum_{j=0}^{D-1}  \ket{j}\bra{j+1}$ is defined via the copying and deleting maps:
\begin{center}
$\delta_X =\left( \begin{array}{c}
\mathbb{I}_{D\times D}\\
P_1(\mathbb{I}_{D\times D})\\
...\\
P_{D-1}(\mathbb{I}_{D\times D})  \end{array} \right)$  and $\epsilon_Z= \frac{1}{\sqrt{D}}(1,0, ..., 0)$.\\
\end{center}
 
Where $\delta_X$ is a $D^2 \times D$ matrix composed of D matrix blocks $\mathbb{I}_{D\times D}$ and $P_j(\mathbb{I}_{D\times D})$ (j=1,2, ... D-1), which are $D \times D$ matrices corresponding to the identity matrix $\mathbb{I}_{D\times D}$, with all its rows permuted to the right by j. 

Unbiased points for the red observable structure are in the form: 
$\ket{\{\alpha_1, \alpha_2, ..., \alpha_{D-1} \}_X}=\ket{+_0}+\sum_{k=1}^{D-1}e^{i \alpha_k}\ket{+_k}=\frac{1}{\sqrt{D}}(c_0\ket{0}+\sum_{j=1}^{D-1} c_j \ket{j})$, where $\ket{+_k}$ are the D eigenvectors of the X matrix. Therefore, the phase group consists of matrices of the form:
\begin{equation}\label{lambx}
\Lambda_X(\alpha_1, \alpha_2, ..., \alpha_{D-1})=\frac{1}{D}\left( \begin{array}{cccccc}
c_0 & c_{D-1} & c_{D-2} & ... & c_2 & c_1  \\
c_1 & c_0 & c_{D-1} & ... & c_3 & c_2 \\
c_2 & c_1 & c_0 & ... & c_4 & c_3 \\
... & ... & ... & ... & ... & ... \\
c_{D-1} & c_{D-2} & c_{D-3} & ... & c_1 & c_0 \end{array} \right)
\end{equation}  

Which can be shown to be unitary and which satisfy: 
\begin{equation}
\Lambda_X(\beta_1, \beta_2, ..., \beta_{D-1}) \circ \Lambda_X(\alpha_1, \alpha_2, ..., \alpha_{D-1}) =\Lambda_X(\alpha_1+\beta_1, \alpha_2+\beta_2, ..., \alpha_{D-1}+\beta_{D-1})
\end{equation}

Therefore the phase group for the red observable $\Pi_X$ and the group of unbiased points $(U_X, \odot_X)$ are both the \textit{\textbf{D-torus group}}, corresponding to the direct product of D circle groups $S^1 \times ...\times S^1$. 

The red part of the ZX calculus for qudits follows from the generalized red spider law.

Note that the red and green observable structures do not induce the same compact structure since: 
\begin{equation}
\eta_Z=\delta_Z \circ \epsilon_Z^{\dagger} \neq \delta_X \circ \epsilon_X^{\dagger}=\eta_X
\end{equation}
The classical points of the green Z observable are \begin{tikzpicture}[scale=0.5]
	\begin{pgfonlayer}{nodelayer}
		\node [style=rn] (0) at (0, 0.5) {};
		\node [style=none] (1) at (0, -0.5) {};
		\node [style=none] (2) at (0.5, 0.5) {k};
	\end{pgfonlayer}
	\begin{pgfonlayer}{edgelayer}
		\draw (0) to (1.center);
	\end{pgfonlayer}
\end{tikzpicture}, where k corresponds to phase values $\{ \alpha_1, ..., \alpha_{D-1} \}$, $\{ \beta_1, ..., \beta_{D-1} \}, ...$ such that the red unbiased points $\ket{\{\alpha_1, ..., \alpha_{D-1} \}_X}$ , $\ket{\{ \beta_1, ..., \beta_{D-1} \}_X \}},...$  are the classical points (eigenvectors) $\ket{0}, \ket{1}, ..., \ket{D-1}$ of the green observable Z. By theorem A5, these D points form an abelian subgroup of the D-torus group $(U_X, \odot_X)$, where $\ket{0}$ is the identity.

Similarly, the classical points of the red X observable are \begin{tikzpicture}[scale=0.5]
	\begin{pgfonlayer}{nodelayer}
		\node [style=gn] (0) at (0, 0.5) {};
		\node [style=none] (1) at (0, -0.5) {};
		\node [style=none] (2) at (0.5, 0.5) {k};
	\end{pgfonlayer}
	\begin{pgfonlayer}{edgelayer}
		\draw (0) to (1.center);
	\end{pgfonlayer}
\end{tikzpicture}, where k corresponds to phase values $\{ \alpha_1, ..., \alpha_{D-1} \}$, $\{ \beta_1, ..., \beta_{D-1} \}, ...$ such that the green unbiased points $\ket{\{\alpha_1, ..., \alpha_{D-1} \}_Z}$ ,$\ket{\{ \beta_1, ..., \beta_{D-1} \}_Z},...$ correspond to the classical points (eigenvectors) $\ket{+_0}, \ket{+_1}, ..., \ket{+_{D-1}}$ of the red observable X. By theorem A5, these D points form an abelian subgroup of the D-torus group $(U_Z, \odot_Z)$, where $\ket{+_0}$ is the identity.

Therefore, the red and green observable structures are a closed pair of coherent observable structures. The (D), (B1) and (B2) rules then follow as before. The (K1) rule becomes:
\begin{center}
\begin{tikzpicture}[scale=0.6]
	\begin{pgfonlayer}{nodelayer}
		\node [style=rn] (0) at (-6, 0.5) {};
		\node [style=gn] (1) at (-6, -0.5) {};
		\node [style=none] (2) at (-6, 1.5) {};
		\node [style=none] (3) at (-7, -1.5) {};
		\node [style=none] (4) at (-5, -1.5) {};
		\node [style=gn] (5) at (-2, 0.5) {};
		\node [style=rn] (6) at (-3, -0.5) {};
		\node [style=rn] (7) at (-1, -0.5) {};
		\node [style=none] (8) at (-2, 1.5) {};
		\node [style=none] (9) at (-3, -1.5) {};
		\node [style=none] (10) at (-1, -1.5) {};
		\node [style=none] (11) at (-4, -0) {=};
		\node [style=none] (12) at (3.5, -0) {=};
		\node [style=none] (13) at (2.5, -1.5) {};
		\node [style=none] (14) at (1.5, 1.5) {};
		\node [style=none] (15) at (6.5, -1.5) {};
		\node [style=none] (16) at (4.5, -1.5) {};
		\node [style=none] (17) at (0.5, -1.5) {};
		\node [style=none] (18) at (5.5, 1.5) {};
		\node [style=none] (19) at (0, -0) {;};
		\node [style=gn] (20) at (1.5, 0.5) {};
		\node [style=rn] (21) at (1.5, -0.5) {};
		\node [style=rn] (22) at (5.5, 0.5) {};
		\node [style=gn] (23) at (4.5, -0.5) {};
		\node [style=gn] (24) at (6.5, -0.5) {};
		\node [style=none] (25) at (-5.5, 0.5) {k};
		\node [style=none] (26) at (-2.5, -0.5) {k};
		\node [style=none] (27) at (-0.5, -0.5) {k};
		\node [style=none] (28) at (2, -0.5) {k};
		\node [style=none] (29) at (6, 0.5) {k};
		\node [style=none] (30) at (9.5, -0) {(K1)};
	\end{pgfonlayer}
	\begin{pgfonlayer}{edgelayer}
		\draw (2.center) to (0);
		\draw (0) to (1);
		\draw (1) to (4.center);
		\draw (1) to (3.center);
		\draw (8.center) to (5);
		\draw (5) to (7);
		\draw (7) to (10.center);
		\draw (6) to (9.center);
		\draw (6) to (5);
		\draw (14.center) to (20);
		\draw (20) to (21);
		\draw (21) to (17.center);
		\draw (21) to (13.center);
		\draw (18.center) to (22);
		\draw (24) to (22);
		\draw (24) to (15.center);
		\draw (23) to (16.center);
		\draw (23) to (22);
	\end{pgfonlayer}
\end{tikzpicture}
\end{center}

There are $2D-2$ equations in (K1), one equation for each of the D classical points of each colour (except the (0,0,..., 0), phaseless points of each colour). Note that if you add to (K1) the case where k corresponds to the (0,0,..., 0), phaseless points of each colour then the rule (B1) follow as a special case of (K1).
 
We obtain the (K2) rule, by calculating the action of the three non trivial red (or green) maps of either colour, corresponding to classical points in $K_Z$ (or $K_X$), on the unbiased green (or red) points in $U_Z$ (or $U_X$). One can then see that the (K2) rule is:

\begin{center}
\begin{tikzpicture}[scale=0.92]
	\begin{pgfonlayer}{nodelayer}
		\node [style=gn] (0) at (-3.5, 2.5) {};
		\node [style=rn] (1) at (1, 2.5) {};
		\node [style=none] (2) at (0, 2) {=};
		\node [style=none] (3) at (-3.5, 3.5) {};
		\node [style=none] (4) at (-3.5, 0.5) {};
		\node [style=none] (5) at (1, 0.5) {};
		\node [style=none] (6) at (1, 3.5) {};
		\node [style=rn] (7) at (-3.5, 1.5) {};
		\node [style=gn] (8) at (1, 1.5) {};
		\node [style=none] (9) at (-2, 2.5) {$\alpha_1,\alpha_2,...,\alpha_{D-1}$};
		\node [style=none] (10) at (-2.75, 1.5) {k};
		\node [style=none] (11) at (1.75, 2.5) {k};
		\node [style=none] (12) at (-3.5, -0.5) {};
		\node [style=gn] (13) at (-3.5, -2.5) {};
		\node [style=none] (14) at (-3.5, -3.5) {};
		\node [style=none] (15) at (0, -2) {=};
		\node [style=none] (16) at (-2, -1.5) {$\alpha_1,\alpha_2,...,\alpha_{D-1}$};
		\node [style=rn] (17) at (1, -2.5) {};
		\node [style=none] (18) at (1, -3.5) {};
		\node [style=none] (19) at (4.5, -2.5) {$\alpha_{k+1}-\alpha_{k},\alpha_{k+2}-\alpha_{k},...,\alpha_{D-1}-\alpha_k, $};
		\node [style=gn] (20) at (1, -1.5) {};
		\node [style=none] (21) at (1, -0.5) {};
		\node [style=rn] (22) at (-3.5, -1.5) {};
		\node [style=none] (23) at (-2.75, -2.5) {k};
		\node [style=none] (24) at (1.75, -1.5) {k};
		\node [style=none] (25) at (10, -0) {(K2)};
		\node [style=none] (26) at (4.5, -3) {$-\alpha_k,\alpha_1-\alpha_{k}, ..., \alpha_{k-1}-\alpha_{k} $};
		\node [style=none] (27) at (4.5, 1) {$-\alpha_k,\alpha_1-\alpha_{k}, ..., \alpha_{k-1}-\alpha_{k} $};
		\node [style=none] (28) at (4.5, 1.5) {$\alpha_{k+1}-\alpha_{k},\alpha_{k+2}-\alpha_{k},...,\alpha_{D-1}-\alpha_k, $};
	\end{pgfonlayer}
	\begin{pgfonlayer}{edgelayer}
		\draw (3.center) to (0);
		\draw (0) to (7);
		\draw (7) to (4.center);
		\draw (6.center) to (1);
		\draw (1) to (8);
		\draw (8) to (5.center);
		\draw (12.center) to (22);
		\draw (22) to (13);
		\draw (13) to (14.center);
		\draw (21.center) to (20);
		\draw (20) to (17);
		\draw (17) to (18.center);
	\end{pgfonlayer}
\end{tikzpicture}
\end{center}

There are 2D-2 equations in (K2) corresponding to the the D phase maps \begin{tikzpicture}[scale=0.7]
	\begin{pgfonlayer}{nodelayer}
		\node [style=none] (0) at (0, 0.75) {};
		\node [style=rn] (1) at (0, -0) {};
		\node [style=none] (2) at (0, -0.75) {};
		\node [style=none] (3) at (0.5, -0) {k};
	\end{pgfonlayer}
	\begin{pgfonlayer}{edgelayer}
		\draw (1) to (2.center);
		\draw (0.center) to (1);
	\end{pgfonlayer}
\end{tikzpicture}
associated to the D classical points for Z and the D phase maps \begin{tikzpicture}[scale=0.7]
	\begin{pgfonlayer}{nodelayer}
		\node [style=none] (0) at (0, 0.75) {};
		\node [style=gn] (1) at (0, -0) {};
		\node [style=none] (2) at (0, -0.75) {};
		\node [style=none] (3) at (0.5, -0) {k};
	\end{pgfonlayer}
	\begin{pgfonlayer}{edgelayer}
		\draw (1) to (2.center);
		\draw (0.center) to (1);
	\end{pgfonlayer}
\end{tikzpicture}
associated to the D classical points for X (except the (0,0,..., 0) phaseless maps for each colour).

For clarity, we illustrate this rule for the case of qudits of dimension four. This requires us to calculate the action of $K_Z$ on $U_Z$:
\footnotesize 
\begin{equation}
\Lambda^{X}(\ket{\{\frac{ \pi}{2}, \pi ,\frac{3 \pi}{2} \}_X}) (\ket{ \{\alpha_1, \alpha_2,\alpha_3 \}_Z})=\left( \begin{array}{cccc}
0 & 1 & 0 & 0  \\
0 & 0 & 1 & 0 \\
0 & 0 & 0 & 1 \\
1 & 0 & 0  & 0  \end{array} \right) \left( \begin{array}{c}
1  \\
e^{i \alpha_1}  \\
e^{i \alpha_2}   \\
e^{i \alpha_3}  \end{array} \right)= e^{i \alpha_1} \left( \begin{array}{c}
1  \\
e^{i (\alpha_2-\alpha_1)}  \\
e^{i (\alpha_3-\alpha_1)}  \\
e^{i (-\alpha_1)}   \end{array} \right)=(\ket{ \{\alpha_2-\alpha_1,\alpha_3-\alpha_1, -\alpha_1 \}_Z})
\end{equation}
\begin{equation}
\Lambda^{X}(\ket{\{ \pi, 0, \pi \}_X}) (\ket{ \{\alpha_1, \alpha_2,\alpha_3 \}_Z})=\left( \begin{array}{cccc}
0 & 0 & 1 & 0  \\
0 & 0 & 0 & 1 \\
1 & 0 & 0 & 0 \\
0 & 1 & 0 & 0  \end{array} \right) \left( \begin{array}{c}
1  \\
e^{i \alpha_1}  \\
e^{i \alpha_2}   \\
e^{i \alpha_3}  \end{array} \right)= e^{i \alpha_2} \left( \begin{array}{c}
1  \\
e^{i (\alpha_3-\alpha_2)}  \\
e^{i (-\alpha_2)}  \\
e^{i (\alpha_1-\alpha_2)}   \end{array} \right)=(\ket{ \{\alpha_3-\alpha_2,-\alpha_2, \alpha_1-\alpha_2 \}_Z})
\end{equation}
\begin{equation}
\Lambda^{X}(\ket{\{\frac{3 \pi}{2}, \pi, \frac{ \pi}{2} \}_X}) (\ket{ \{\alpha_1, \alpha_2,\alpha_3 \}_Z})=\left( \begin{array}{cccc}
0 & 0 & 0 & 1  \\
1 & 0 & 0 & 0 \\
0 & 1 & 0 & 0 \\
0 & 0 & 1 & 0  \end{array} \right) \left( \begin{array}{c}
1  \\
e^{i \alpha_1}  \\
e^{i \alpha_2}   \\
e^{i \alpha_3}  \end{array} \right)= e^{i \alpha_3} \left( \begin{array}{c}
1  \\
e^{i (-\alpha_3)}  \\
e^{i (\alpha_1-\alpha_3)}  \\
e^{i (\alpha_2-\alpha_3)}   \end{array} \right)=(\ket{ \{-\alpha_3,\alpha_1-\alpha_3, \alpha_2-\alpha_3 \}_Z})
\end{equation}
\normalsize
The action of $K_X$ on $U_X$ is exactly dual to this.

The last rule will correspond to the definition of the Fourier gate $F=\frac{1}{\sqrt{D}}\sum_{j,k=0}^{D-1}\eta^{jk}\ket{j}\bra{k}$ in the calculus.
In general, one can show that:
\begin{equation}
(F \otimes F) \circ \delta_Z \circ F^{\dagger}= \delta_X
\end{equation}
and:
\begin{equation}
F(\ket{\{\alpha_1,\alpha_2,... \}_Z})=\ket{\{\alpha_1,\alpha_2,... \}_X}
\end{equation}
where F is the unitary Fourier matrix. This holds for all dimensions and allows us to introduce the Fourier gate in the qudit ZX calculus in much the same way as the Hadamard matrix was introduced in the qubit ZX calculus\cite{Du08}, except that the Fourier gate corresponds to box with a vertical (involutive) asymmetry. This gives us the (F) rules of the qudit ZX calculus:

\begin{center}
\begin{tikzpicture}[scale=0.7]
	\begin{pgfonlayer}{nodelayer}
		\node [style=rn] (0) at (-4.5, 2.5) {};
		\node [style=none] (1) at (-4.5, 0.4999999) {...};
		\node [style=none] (2) at (-4.5, 4.5) {...};
		\node [style=none] (3) at (7.499999, 2.5) {(F1)};
		\node [style=none] (4) at (3, -3) {};
		\node [style=none] (5) at (3, -0.5000002) {};
		\node [style=none] (6) at (7.499999, -1.75) {(F2)};
		\node [style=gn] (7) at (-4.5, 2.5) {};
		\node [style=none] (8) at (-3, -3) {};
		\node [style=none] (9) at (-3, -0.5000002) {};
		\node [style=none] (10) at (-1.5, -1.75) {=};
		\node [style=none] (11) at (-2.5, 2.5) {$\alpha_1,\alpha_2, ..., \alpha_{D-1} $};
		\node [style=Had] (12) at (-3, -1.25) {$F^{\dagger}$};
		\node [style=Had] (13) at (-3, -2.25) {$F$};
		\node [style=none] (14) at (0, -0.5000002) {};
		\node [style=Had] (15) at (0, -2.25) {$F^{\dagger}$};
		\node [style=none] (16) at (0, -3) {};
		\node [style=Had] (17) at (0, -1.25) {$F$};
		\node [style=none] (18) at (1.5, -1.75) {=};
		\node [style=Had] (19) at (-6, 3.5) {$F^{\dagger}$};
		\node [style=Had] (20) at (-5, 3.5) {$F^{\dagger}$};
		\node [style=Had] (21) at (-4, 3.5) {$F^{\dagger}$};
		\node [style=Had] (22) at (-3, 3.5) {$F^{\dagger}$};
		\node [style=Had] (23) at (-6, 1.5) {F};
		\node [style=Had] (24) at (-5, 1.5) {F};
		\node [style=Had] (25) at (-4, 1.5) {F};
		\node [style=Had] (26) at (-3, 1.5) {F};
		\node [style=none] (27) at (-6, 4.5) {};
		\node [style=none] (28) at (-5.25, 4.5) {};
		\node [style=none] (29) at (-3.75, 4.5) {};
		\node [style=none] (30) at (-3, 4.5) {};
		\node [style=none] (31) at (-3, 0.4999999) {};
		\node [style=none] (32) at (-3.75, 0.4999999) {};
		\node [style=none] (33) at (-5.25, 0.4999999) {};
		\node [style=none] (34) at (-6, 0.4999999) {};
		\node [style=none] (35) at (2.5, 4.5) {...};
		\node [style=none] (36) at (0.9999997, 4.5) {};
		\node [style=none] (37) at (1.75, 4.5) {};
		\node [style=none] (38) at (4, 4.5) {};
		\node [style=none] (39) at (4, 0.4999999) {};
		\node [style=none] (40) at (1.75, 0.4999999) {};
		\node [style=none] (41) at (2.5, 0.4999999) {...};
		\node [style=none] (42) at (4.5, 2.5) {$\alpha_1,\alpha_2, ..., \alpha_{D-1} $};
		\node [style=none] (43) at (3.25, 4.5) {};
		\node [style=none] (44) at (3.25, 0.4999999) {};
		\node [style=none] (45) at (0.9999997, 0.4999999) {};
		\node [style=none] (46) at (0, 2.5) {=};
		\node [style=rn] (47) at (2.5, 2.5) {};
	\end{pgfonlayer}
	\begin{pgfonlayer}{edgelayer}
		\draw [style=none] (5.center) to (4.center);
		\draw (9.center) to (12);
		\draw (12) to (13);
		\draw (13) to (8.center);
		\draw (14.center) to (17);
		\draw (17) to (15);
		\draw (15) to (16.center);
		\draw (0) to (21);
		\draw (0) to (22);
		\draw (22) to (30.center);
		\draw (21) to (29.center);
		\draw (28.center) to (20);
		\draw (20) to (0);
		\draw (0) to (19);
		\draw (19) to (27.center);
		\draw (0) to (23);
		\draw (23) to (34.center);
		\draw (0) to (24);
		\draw (24) to (33.center);
		\draw (0) to (25);
		\draw (25) to (32.center);
		\draw (0) to (26);
		\draw (26) to (31.center);
		\draw (36.center) to (47);
		\draw (37.center) to (47);
		\draw (43.center) to (47);
		\draw (47) to (38.center);
		\draw (47) to (45.center);
		\draw (40.center) to (47);
		\draw (47) to (44.center);
		\draw (47) to (39.center);
	\end{pgfonlayer}
\end{tikzpicture}
\end{center}
Therefore, we have justified all the rules of the qudit ZX calculus from the algebraic properties of the Z and X observables and of the Fourier map.
This construction, together with Theorem A1, shows that qudit ZX calculus is \textbf{sound} for quantum mechanics.

\section{Qudit stabilizer quantum mechanics}

We describe the generalization of qubit stabilizer quantum mechanics \cite{Got97} to quantum systems of dimension D, where D can be higher\cite{Gott98} than 2. 
Stabilizer states are eigenstates with eigenvalue 1 of each operator in a subgroup of the generalized Pauli group of operators acting on the Hilbert space of n qudits:
 
\begin{equation}
\mathcal{P}_{D,n}:= \{ \sqrt{\eta}^{\lambda} g_1 \otimes ... \otimes g_n : \eta = e^{\frac{2 \pi i}{D}} \wedge \lambda \in \mathbb{Z}_{2D} \} 
\end{equation}
with: $g_k= X^{x_k} Z^{z_k} $ and $ x_k,z_k \in \mathbb{Z}_D;   \forall k \in \{1, ..., n \}$. Note that sums and multiplication are all modulo D and $\mathbb{Z}_D$ are integers modulo D.
 
The single qudit Z and X operators are: 
\begin{equation}
Z=\sum_{j=0}^{D-1} \eta^j \ket{j}\bra{j} \; \; \; \text{and}  \; \; \; X=\sum_{j=0}^{D-1} \ket{j}\bra{j+1}
\end{equation}

One can easily see that: $XZ=\eta ZX$ and $Z^D=X^D=\mathbb{I}$. \\

The generalized Clifford group on n qudits consists of the unitary operations that leave Pauli operators invariant under conjugation:
\begin{equation}
C_n := \{ U : U g U^{\dagger} \in \mathcal{P}_{D,n}, \forall g \in \mathcal{P}_{D,n} \}
\end{equation}
The following gates are generalizations of standard qubit gates to higher dimensions \cite{Ghe11}. \\

The generalization of the Hadamard gate is the Fourier gate: $F:= \frac{1}{\sqrt{D}} \sum_{j,k=0}^{D-1} \eta^{jk} \ket{j}\bra{k}$. Another important set of qudit gates are the multiplicative gates: $S_q:=  \sum_{j=0}^{D-1}  \ket{j}\bra{jq}$, where $q \in \mathbb{Z}_D$ such that $\exists \bar{q}\in \mathbb{Z}_D$ with $q\bar{q}=1$.

We define the qudit controlled NOT and controlled phase gates between control qudit \textbf{a} and target qudit \textbf{b} as:  
\begin{equation}
CNOT_{a,b}:=\sum_{j,k=0}^{D-1}  \ket{k}\bra{j}_a \otimes \ket{k}\bra{k+j}_b \; \; \; \; \text{and}  \; \; \; \; CP_{a,b}:=\sum_{j,k=0}^{D-1}  \eta^{jk} \ket{j}\bra{j}_a \otimes \ket{k}\bra{k}_b 
\end{equation} 
The swap gate is: $SWAP_{a,b}:=\sum_{j,k=0}^{D-1}  \ket{k}\bra{j}_a \otimes \ket{j}\bra{k}_b $. Note that the SWAP gate can be decomposed as: 
\begin{equation}
SWAP_{a,b}=CNOT_{a,b} CNOT_{b,a}^{\dagger} CNOT_{a,b}(F_a^2 \otimes \mathbb{I}_b)
\end{equation}
Similarly to the qubit case, the controlled phase gate can be decomposed as:  
\begin{equation}
CP_{a,b}=(\mathbb{I}_a \otimes F_b)^{\dagger} CNOT_{a,b} (\mathbb{I}_a \otimes F_b)
\end{equation} 
The generalized Clifford group is generated \cite{Ghe11, Hos05} by the set of three gates: $\{ F, S_q, CNOT_{a,b} \}$.

Stabilizer quantum mechanics \cite{Got97} includes state preparations in the computational basis $\{ \ket{0}, \ket{1}, \ket{2}, ...\}$, generalized Clifford unitaries and measurements of observables in the generalized Pauli group. In addition to its foundational importance, the theory of qudit stabilizer quantum mechanics plays a key role in quantum information theory, in quantum key distribution and in quantum error correction.

Extending the Gottesman-Knill theorem shows that qudit stabilizer quantum mechanics can be efficiently simulated by a classical computer. Indeed, a group of order K has at most log(K) generators therefore the qudit stabilizer group can be compactly described using the group generators. One can show that if D is prime then any n-dimensional stabilizer group can be described using at most n generators \cite{Gott98}. In composite dimensions one can have more than n generators but no more than 2n \cite{Ghe11}.

\section{Spekkens toy theory in higher dimensions}

Previous work in quantum foundations \cite{Spek07, Rud12, Scsp12} has shown that considering a classical statistical theory together with a fundamental restriction on the allowed statistical distributions over phase space allows one to reproduce a large part of operational quantum mechanics. We will now introduce some of this work for physical systems with discrete degrees of freedom \cite{Scsp12}. We call the theory described here \textbf{Spekkens-Schreiber toy theory for dits}. 

Let phase space $\Omega=(\mathbb{Z}_d)^{2n}$ consist of a set of points (ontic states) $m \equiv (x_1, p_1, ..., x_n, p_n) \in \Omega$.

We can then define functionals on phase space $F: \Omega \rightarrow \mathbb{Z}_d$ and a Poisson bracket of functionals:
\begin{equation}
\{F, G \}(m):= \sum_{j=1}^{n}(F[m+e_{x_j}]-F[m])(G[m+e_{p_j}]-G[m])-(F[m+e_{p_j}]-F[m])(G[m+e_{x_j}]-G[m])
\end{equation} 
where $e_{x_j}$ and $e_{p_j}$ have a 1 in position $x_j$ and $p_j$ respectively and zeros everywhere else. \\
We define canonical variables as the linear functionals:
\begin{equation}
F=a_1 X_1+b_1 P_1+ ... + a_n X_n +b_n P_n
\end{equation}
\begin{equation}
G=c_1 X_1+d_1 P_1+ ... + c_n X_n +d_n P_n
\end{equation}

where $X_k(m)=x_k$, $P_k(m)=p_k$ and $a_j, b_j, c_j, d_j \in \mathbb{Z}_d$, $\forall j \in \{1, ..., n \}$.

These form the dual space $\Omega^{\star}\equiv (\mathbb{Z}_d)^{2n}$ such that: $F=(a_1, b_1, ..., a_n, b_n), G=(c_1,d_1, ..., c_n, d_n) \in \Omega^{\star}$.
We can then write the Poisson bracket of canonical variables as a sympectic inner product of vectors:
\begin{equation}
\{F, G \}(m)= \sum_{j=1}^{n}(a_j d_j-b_j c_j)= F^{T} J G 
\end{equation}
where:
\begin{equation}
J=\bigoplus_{k=1}^{n} \left( \begin{array}{cc}
0 & -1  \\
1 & 0 
 \end{array} \right)
 \end{equation}

We then define the principle of \textbf{classical complementarity} in the following way: an observer can only have knowledge of the values of a commuting set of canonical variables (whose Poisson brackets all vanish) and is maximally ignorant otherwise.

The Spekkens-Schreiber toy theory for dits can then be described in the following way:

(a) \textit{Valid epistemic states} are specified by isotropic subspaces $V\subseteq \Omega^{\star}$, such that $\{F, G\}=0; \; \forall F,G \in V $, together with a valuation vector $v:V \rightarrow \mathbb{Z}_d$ ($v \in V^{\star}$) such that: $v(F)=F^{T} v; \; \forall F\in V$. Therefore, V specifies which set of canonical variables are known and v describes what is known about them (in analogy with the commuting set of eigenoperators of the quantum state, together with their eigenvalues). 

Epistemic states can also be characterized by a probability distribution over phase space $\Omega$. We can define the orthogonal complement of V as:
\begin{equation}
V^{\bot}:=\{m \in \Omega \vert P_V m=0 \}
\end{equation}
where $P_V$ is the projector onto V. Note that the phase space points $m \in \Omega$ which are consistent with an epistemic state associated to the isotropic subspace V and valuation vector v are those which satisfy:
\begin{equation}
F^{T}m=F^{T}v, \; \forall F \in V
\end{equation}

Therefore, the probability distribution for the epistemic state associated to the isotropic subspace V and valuation vector v is:
$p_{V,v}: \Omega \rightarrow [0,1]$ such that:
\begin{equation}
p_{V,v}(m)= \frac{1}{\vert V^{\bot} \vert} \delta_{V^{\bot}+v}(m)
\end{equation}
where $\vert V^{\bot} \vert$ is the cardinality of $V^{\bot}$ and $\delta_{V^{\bot}+v}(m)$ is 1 if $m \in V^{\bot}+v$ and zero otherwise.

(b) \textit{Valid reversible transformations} correspond to all the symplectic, affine transformations (analogues of the Clifford operations). These are the phase space maps $C:\Omega \rightarrow \Omega$ such that: $C(m)=S m+a $ where $a \in \Omega$ and $\{Su, Sv\}=\{u,v\}, \forall u,v \in (\mathbb{Z}_d)^{2n}$.

(c) \textit{Valid measurements} are described by sets of indicator functions $\xi_k:\Omega \rightarrow [0,1]$ such that $\sum_k \xi_k=u$ (where u is a function mapping every point of phase space to 1) which correspond to some choice of a set of non-conjugate variables. The outcome probability can then be obtained by:
\begin{equation}
p_k=\sum_{\lambda} \mu(\lambda) \xi_k(\lambda)
\end{equation}
where $\mu(\lambda)$ is the epistemic state.
   
The Spekkens-Schreiber theory for any number of dits of any dimension can be represented using matrices to describe the valid epistemic states, transformations and measurements. This corresponds to the subcategory of \textbf{FRel} which we described earlier.

Note that Spekkens toy model for bits \cite{Spek07} is a special instance of Spekkens-Schreiber theory for dimension 2 and that the `knowledge balance principle' is superseded by the principle of classical complementarity described above.
 
\end{appendices}

\bibliographystyle{eptcs}
\bibliography{Mresbib}

\end{document}